# Material matters in superconducting qubits

Conal E. Murray*

IBM Quantum

IBM T.J. Watson Research Center, Yorktown Heights, NY 10598  USA

The progress witnessed within the field of quantum computing has been enabled by the identification and understanding of interactions between the state of the quantum bit (qubit) and the materials within its environment.  Beginning with an introduction of the parameters used to differentiate various quantum computing approaches, we discuss the evolution of the key components that comprise superconducting qubits, where the methods of fabrication can play as important a role as the composition in dictating the overall performance.  We describe several mechanisms that are responsible for the relaxation or decoherence of superconducting qubits and the corresponding methods that can be utilized to characterize their influence.  In particular, the effects of dielectric loss and its manifestation through the interaction with two-level systems (TLS) are discussed.  We elaborate on the methods that are employed to quantify dielectric loss through the modeling of energy flowing through the surrounding dielectric materials, which can include contributions due to both intrinsic TLS and extrinsic aspects, such as those generated by processing.  The resulting analyses provide insight into identifying the relative participation of specific sections of qubit designs and refinements in construction that can mitigate their impact on qubit quality factors.  Additional prominent mechanisms that can lead to energy relaxation within qubits are presented along with experimental techniques which assess their importance.  We close by highlighting areas of future research that should be addressed to help facilitating the successful scaling of superconducting quantum computing.

*E-mail:  conal@us.ibm.com

**OUTLINE**

1. Introduction
    1. Background
    2. Quantum computing approaches
    3. Key parameters
2. Dielectric Loss
    1. Two-level systems
    2. Loss tangent
    3. Extrinsic effects
    4. Spectroscopic Characterization
    5. Summary
3. Participation
    1. Key interfaces
    2. Substrate
    3. Modeling approaches
    4. Effects of substrate geometry (trenching)
    5. Comparison to experimental measurements
    6. Summary
4. Metallization
    1. Composition
    2. Texture
    3. Processing effects
    4. Inductive loss
    5. Summary
5. Tunnel junctions
    1. Composition
    2. Formation
    3. Characterization
    4. Summary
6. Radiation effects
    1. Quasiparticles
    2. Characterization
    3. Radiation Loss
    4. Summary
7. Decoherence issues
    1. Dephasing
    2. Noise
    3. Summary
8. Conclusions and future challenges

# 1. Introduction

## 1.1 Background

One of the key benefits heralded by the promise of quantum computing, as posited by Richard Feynman [1], is the ability to treat problems associated with quantum mechanics that cannot be simulated using classical computing schemes. Its realization relies on the stochastic and sometimes counterintuitive interactions that govern quantum effects [2-4]. Despite the complexity involved in its operation, tremendous advances in the performance of quantum bits (qubits), the building blocks of quantum computing, have been achieved over the past two decades [5],[6]. Such accomplishments provide the opportunity to study problems which require finding an optimal solution among a large and complex landscape that would be difficult or intractable to span even with the use of massively parallel computing clusters. In contrast to classical, binary bits which can assume one of two states ("0" or "1"), qubits can be composed of a linear superposition of ground ($|0\rangle$) and excited states ($|1\rangle$). The property of entanglement, which dictates nonlocal interactions among qubits, in concert with superposition is responsible for enabling this rapid growth in solution space that can be interrogated. Although current implementations employ a small number of qubits (less than 100) relative to that anticipated to be necessary in delivering a significant advantage over conventional algorithms, near-term quantum computing has already been successfully applied to investigations in the fields of chemistry [7]-[11], biology [12], solid state [13]-[15] and particle physics [16], machine learning [17],[18] and finance [19],[20]. However, in the absence of error correction, the fragile nature of the qubit state requires establishing a balance between its interactions and its isolation from many forms of ambient energy emanating from the environment to avoid a loss of coherence.

## 1.2 Quantum computing approaches

Several platforms [21] have been demonstrated, and others proposed, on which quantum computing can be performed, including photons [22],[23], solid-state, spin-based systems [24],[25], defects [26] and impurities [27] in dielectric materials, Majorana fermions [28], trapped ions [29],[30] and superconducting electronics [31],[32]. In this article, we focus on the last category which itself contains a variety of possible manifestations [33]. In superconducting qubits, Josephson junctions and capacitive elements comprise anharmonic resonators that act as artificial atoms, in which the two lowest energy states correspond to the ground ($|0\rangle$) and excited ($|1\rangle$) states. The transition frequency between these two states, $f_{10}$, often resides in the microwave regime. Different assemblages of these elements have been implemented to form qubits that encompass different regimes of operational space with respect to the conjugate variables of charge and flux across the junction [6],[34]-[36]. A quantity useful in categorizing these designs involves the ratio of the junction energy, $E_J = \hbar^2/(4e^2 L_J)$, to the charging energy, $E_C = e^2/(2C_Q)$, where $L_J$ refers to the Josephson junction inductance and $C_Q$ the total qubit capacitance, composed of the Josephson junction and shunting capacitors. For example, one of the first iterations of qubits, the Cooper pair box [37], possesses an $E_J/E_C$ ratio close to 1 while that of a transmon (transverse plasma oscillation) qubit is greater than 30. Since sensitivity to charge noise, which can dramatically affect the stability of the transition frequency in charge qubits, decreases exponentially as $E_J/E_C$ increases, transmons are much more robust against this decoherence than Cooper pair boxes [34]. **Fig. 1a** depicts a typical fixed-frequency transmon [38] in which one Josephson junction is located between the shunting capacitor ($C_S$) paddles, and capacitively coupled ($C_C$) to a coplanar waveguide (CPW) resonator and its equivalent circuit diagram in **Fig. 1b**. Another important parameter is the $L_J/L$ ratio [6],[35], where L refers to a shunting inductance in parallel with the Josephson junction, in addition to any shunting capacitance. A class of superconducting qubits that exploit increased inductance, either through loops [39],[40] or arrays of Josephson junctions [41],[42], can be tuned through the application of flux and may possess improved protection of the qubit eigenstates which are delocalized with respect to the flux operator [43].

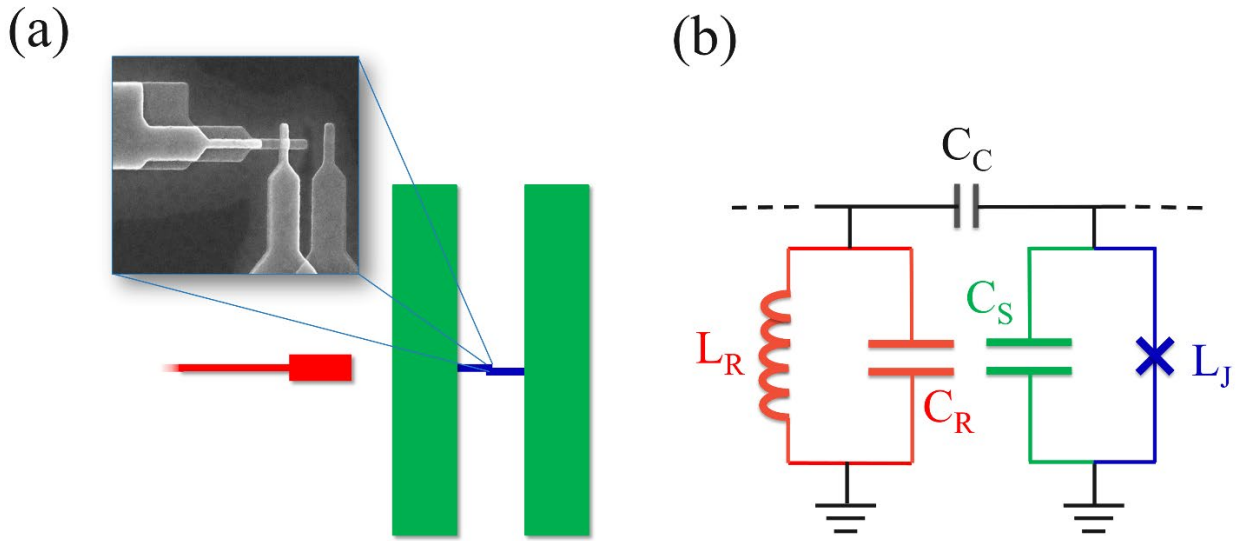

FIG. 1: (a) Top-down schematic of a fixed-frequency transmon qubit, possessing shunting capacitors (green) with capacitance, $C_S$, in parallel with a single Josephson junction (blue) possessing a nonlinear inductance, $L_J$. The inset contains an SEM image depicting the Josephson junction located at the intersection of the two metal electrodes. Communication with the qubit is accomplished through a coplanar waveguide resonator (red), with capacitance $C_R$ and inductance $L_R$, that is capacitively coupled ($C_C$) to the transmon. (b) Equivalent circuit diagram of (a).

A common theme to many of these qubit designs is their fabrication from lithographically patterned superconducting metallization to create the circuit elements necessary for their implementation. Such structures can be coplanar in geometry to form either lumped element capacitors, inductors or much larger features such as CPW resonators and groundplanes on one surface. In addition, the nonlinear inductance generated by Josephson junctions can be generated through passivation of the metal electrodes or direct deposition and etching of thin, insulating layers. Although initial readouts of the qubit state involved direct voltage biasing as a function of microwave frequency [44], qubit control and measurement in the dispersive limit using either three dimensional cavities [45] or two dimensional, coplanar resonators [46] offer better isolation from the external environment. However, future needs associated with qubit scaling may also necessitate the incorporation of three-dimensional architectures that encompass regions within the substrate [47].

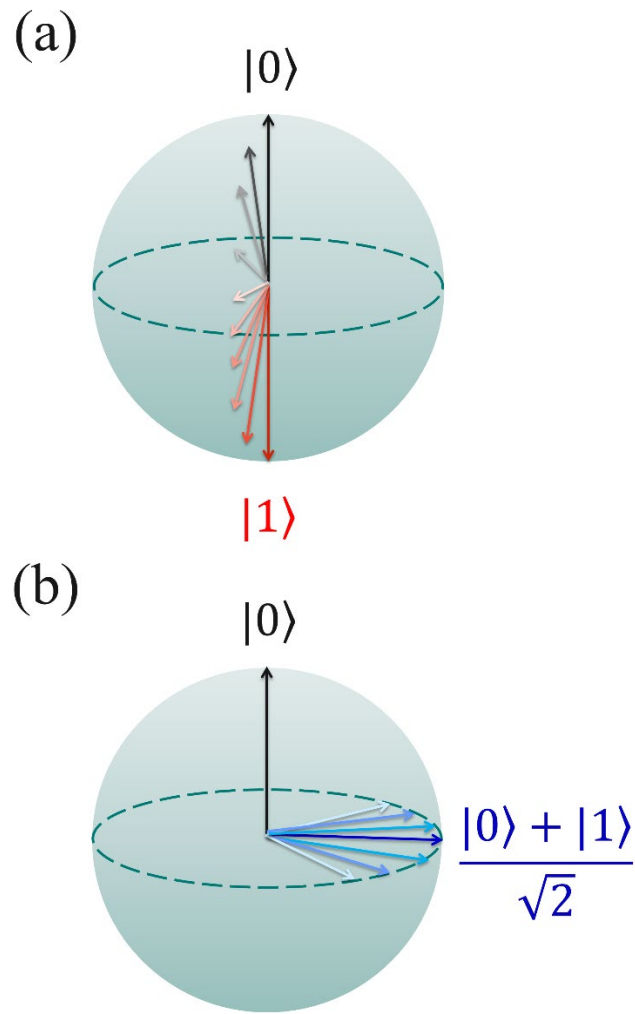

FIG. 2: (a) Bloch-sphere representation of energy relaxation of a qubit, corresponding to $T_1$ decay and (b) dephasing of a qubit which, in combination with relaxation, contributes to the decoherence time, $T_2$.

## 1.3 Key parameters

**Fig. 2a** is a common representation of the relaxation of a qubit using the Bloch sphere from the excited (bottom pole) to ground state (top pole) over a characteristic time constant associated with its exponential decay, $T_1$. Because qubits possess phase as well as amplitude information, its decoherence refers to the unwanted movement of the qubit state in the equatorial plane of the Bloch sphere, as shown in **Fig. 2b** in the case of a superposition composed of $|0\rangle$ and $|1\rangle$ states. The dephasing time, $T_\varphi$, is incorporated with qubit relaxation ($T_1$) in the expression for its coherence time, $T_2$:

$$\frac{1}{T_2} = \frac{1}{2T_1} + \frac{1}{T_\varphi} \quad (1)$$

where the analogs of $T_1$ and $T_2$ to the longitudinal and transverse relaxation processes in NMR spectroscopy are apparent [48]-[50]. The corresponding quality factor, Q, of qubits consists of the product of $T_1$ with the angular frequency, $\omega_1 = 2\pi f_1$, where $f_1$ represents the excited to ground state transition. A similar metric for resonators, the internal quality factor, $Q_i$, can be extracted by fitting the $S_{21}$ transmission across a coplanar feedline coupled to the resonators while sweeping the frequency across a specified range [51],[52], and can be represented by:

$$\frac{1}{Q_t} = \frac{1}{Q_i} + \frac{1}{Q_c} \quad (2)$$

where $Q_t$ and $Q_c$ refer to the total (or loaded) and coupling quality factors, respectively. This simple expression belies the complexity in extracting intrinsic loss in resonators because $Q_i$ is a function of several parameters, such as resonator geometry, measurement temperature and signal power [53].

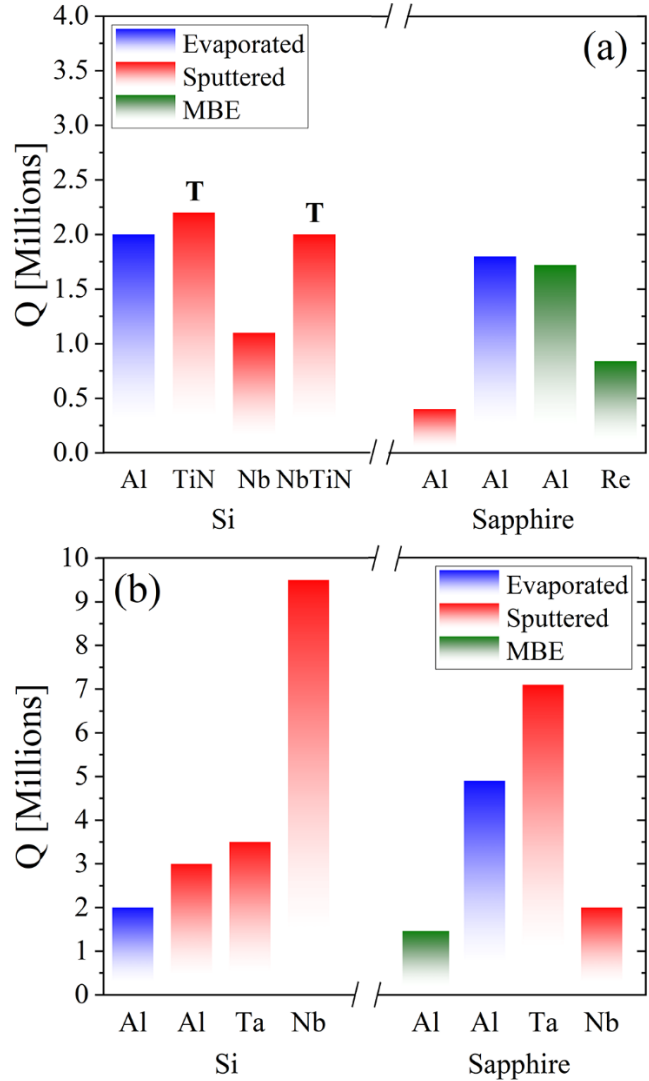

FIG. 3: Comparison of (a) internal quality factors of CPW resonators for a variety of deposition methods of superconducting metallization and substrates: sputtered [51],[63]-[65], evaporated [66],[67] and molecular beam epitaxy (MBE) [51],[68]. (b) Comparison of quality factors of transmon qubits fabricated using different deposition methods and superconducting metallization: sputtered [69],[70],[71] evaporated [72],[73] and MBE [74].

**Figs. 3a and 3b** contain the internal quality factors, $Q_i$, reported in the literature for superconducting CPW resonators and those of transmon qubits measured in the single photon regime below a temperature of 100 millikelvin, where the top of the bars represent the largest average values. Though implementations of solid state qubits exist with Q values that exceed these maxima [54]-[56], these designs employ additional protection of the qubit state through its localization in the potential energy landscape[1]. These figures are meant to present a general survey extracted from various material combinations using similar geometries. In fact, all of the qubits displayed in **Fig. 3b** possess Josephson junctions with an aluminum oxide tunnel barrier and aluminum leads, which could explain the similarity in top quality factors in the range of 5 to 10 million but using different groundplane and shunting capacitor metallization choices (e.g: Al, Nb and Ta). Variations with respect to design and geometry, filtering and shielding within the cryogenic testing apparatus, all contribute to differences among the results. For example, resonators highlighted by 'T' employ deep trenches etched in the underlying dielectric substrate, and exhibit better performance. Although we will elaborate on the effects stemming from these and other mechanisms in the following sections, we can make some general observations at this point. Based on **Fig. 3a**, it appears that the quality factors of CPW's (without trenched substrates) may saturate at approximately 2 million. The fact that qubits demonstrate larger Q values could be a manifestation that CPW resonators are more susceptible to a particular loss mechanism, such as quasiparticles or radiation effects, which may not impact qubits as strongly[2].

---

[1] For example, fluxonium [35],[42] and 0-π qubits [43] have demonstrated $T_1$ times of 1 millisecond and larger.

[2] Recent investigations that report quality factors of $5 \times 10^6$ and better for Nb resonators on Si substrates [57],[58] employ prolonged chemical etching of the metallic oxide layers (see Section 4.3).

However, there are other methods to interrogate qubits, such as those using three-dimensional cavity resonators that exhibit $T_1$ values up to 10 ms [56],[59]-[62], which could help to map out conditions where Purcell loss from CPW resonators influences superconducting qubit quality factors.

This review article is organized as follows: Section 2 will focus on the various aspects of dielectric loss that can impact qubit performance. In Section 3, we describe ways to calculate loss mechanisms within various resonator and qubit designs, followed by the effects pertinent to superconducting metallization and tunneling barriers in Sections 4 and 5, respectively. Issues associated with quasiparticles and radiation loss more generally are detailed in Section 6, followed by decoherence and noise effects in Section 7. We conclude with a summary of the challenges and future opportunities associated with superconducting qubits in Section 8.

## 2. Dielectric loss
### 2.1 Two-level systems

One of the first mechanisms identified as inducing qubit relaxation, dielectric loss occurs by the siphoning of energy from the coherent state of the qubit, leading to a decay to its ground state. Electromagnetic energy emanating from the qubit metallization interacts with imperfections in the surrounding dielectric material, described as two level systems (TLS) in which tunneling between two states separated by an activation barrier, absorbs this energy. According to the Generalized Tunneling Model (GTM), at least two species of TLS, one with an energy difference commensurate with that of the qubit and a sea of lower energy TLS, closer to that of $k_BT$, can lead to decoherence and frequency variations within the qubit, where tunneling through an activation barrier allows the structure to transition to another state [75]. Although the specific species may vary with respect to the host dielectric, there are common traits to the description of TLS behavior, such as a net dipole moment which couples to the incident electric fields. Amorphous materials represent primary candidates for hosting TLS due to their propensity for containing many structural arrangements that are close in aggregate energy. However, it is the nature of the bonding within amorphous structures that dictates their response. For example, defect bond configurations, which can often be found in voids or regions of low density, can more easily couple by tunneling [76],[77] than tetrahedral networks, or highly coordinated atoms [76] formed in amorphous Si [78],[79]. The fact that many amorphous glasses can be described by such a model, in addition to disordered crystals [80], suggests that commonalities exist in their tunneling interactions.

Unfortunately, this broad description makes it difficult to identify specific aspects of individual amorphous materials that may be responsible for TLS. In fact, they are so ubiquitous that the challenges of managing TLS-based effects have been engineered as a platform for quantum computing, either within the Josephson junction [81] or using the dangling bonds present on Si (001) surfaces [82]. Coupling between the TLS and qubit can be so strong that the measured relaxation rate, which normally follows an exponential decrease ($T_1$), can exhibit oscillatory behavior [75]. The large heat capacity observed in amorphous solids has been attributed to TLS [76],[83]. Though predictions of glass formed through infinitesimally slow cooling to an "ideal" state are anticipated to possess similar entropy to that of a crystalline state with fewer TLS [84], measurements conducted on amber glass indicate that the effect of TLS did not diminish in amber glasses after 110 million years of aging while atomic density slightly increased [85]. These seemingly inconsistent observations may be reconciled by considering the presence of impurities that create dipole moments within dielectrics as being responsible.

## 2.2 Loss tangent

Coupling of an arbitrary TLS to the incident electric field is affected by many parameters, such as the relative orientation of its dipole moment, proximity of the TLS frequency of that of the radiation, as well as physical proximity to sections of maximum electric field. For a sufficiently broad distribution of TLS, their response can be represented by a continuum approximation [86]. The dielectric loss tangent, the ratio of the imaginary to real component of the dielectric function in a material, is such a continuum description of an inherently atomistic process. External factors that can influence the impact of TLS on superconducting resonators and qubits include the reduction of their energy absorption, termed saturation, which occurs when the resonance (or Rabi) frequency of the TLS exceeds that dictated by the combination of its own relaxation and decoherence times: $1/\sqrt{T_1^{TLS} T_2^{TLS}}$ [75],[86]. Saturation can be accomplished by increasing the temperature of the ambient or magnitude of the incident power applied to the resonator. Under the assumption of weak electric fields[+], an expression for these factors often takes the form:

---

[+] At higher powers, resonator frequencies can also be impacted by non-resonant TLS due to a larger shift in the effective dielectric constant [75],[90].

$$\tan(\delta) = \tan(\delta_0) \tanh\left(\frac{hf}{2k_BT}\right) / \sqrt{1 + \left(\frac{x}{x_c}\right)^n} \qquad (3)$$

which encompasses the contributions due to qubit or resonator frequency, $f$, temperature, T, and to the ratio of a parameter, x, associated with the input energy that scales at a particular exponent, n. When x explicitly corresponds to energy density [87], power density [88],[89] or photon number [53],[90], n is usually chosen to be 1. For variables more closely linked to measurement parameters, such as voltage or electric field magnitude [91],[92], an exponent of 2 is used [93],[94]. $x_c$ refers to a critical value for the specific parameter, and provides a representation for the reduction in loss tangent from its maximum value, $\tan(\delta_0)$, due to saturation.

From Eq. 3, it is clear that TLS effects are much larger at the cryogenic temperatures associated with typical superconducting qubit operation where, for microwave frequencies, $hf$ is typically 5 to 10 times larger than $k_bT$. The observations of increased quality factor with decreasing CPW resonator frequency has been explained by a decrease in TLS polarization [63]. While theoretical investigations have suggested specific atomic species and configurations that may be responsible [95]-[97], quantification of $\tan(\delta_0)$ for dielectrics necessitates experimental verification through the analysis of oscillator dissipation. The vast majority of these measurements involve resonators, either CPW or lumped element in design, which can allow for the deposition and patterning of a specific material under investigation. Such an approach naturally assumes that the electric field energy mainly interacts with the lossy dielectric, rather than the surrounding materials. Fortunately, the high level of crystalline quality inherent in current sapphire and silicon substrates, due to decades in advancement in growth and refinement processes [98],[99], can minimize their contribution to loss.

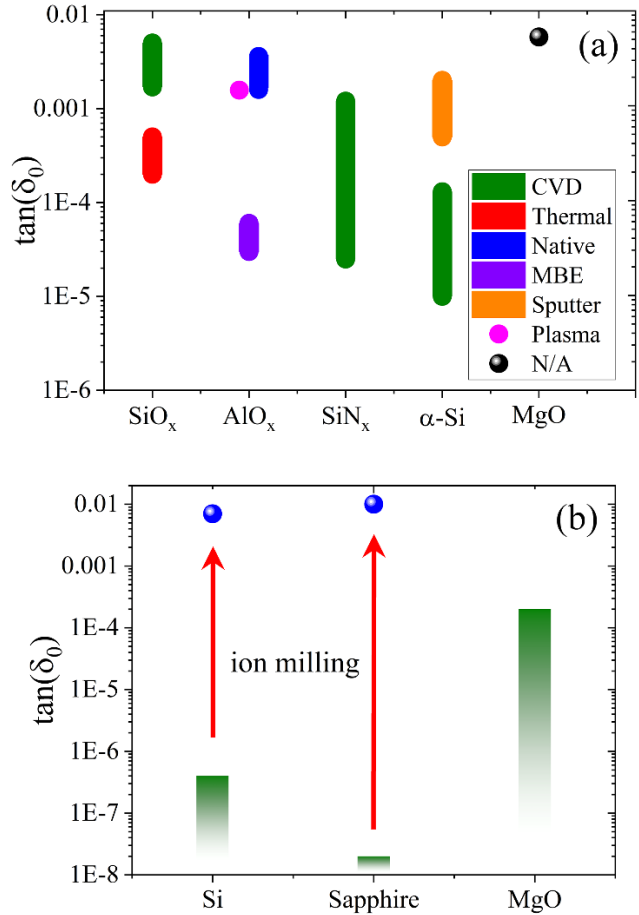

FIG. 4: Comparison of loss tangents extracted from measurements below 100 mK for (a) $SiO_x$ [77],[100],[101], $AlO_x$ [102]-[105], $SiN_x$ [106], amorphous Si [77],[100] and MgO [77] dielectric films, and (b) single crystal substrates [107]-[109]. A wide range of values can be observed due to differences in fabrication and the corresponding impurity levels and defectivity. For substrates, where the bars represent the upper bounds based on resonator and qubit measurements, damage due to argon ion milling is depicted by the blue dots [66],[104].

**Fig. 4a and 4b** illustrates the loss tangents of several types of dielectric films and single crystal substrates, respectively, measured below 100 mK. We focus on these temperatures and low incident power to mitigate TLS saturation, which often leads to much higher apparent quality factors in superconducting resonators. Oxide films possess loss tangents in the range of mid $10^{-4}$ to mid $10^{-3}$, where thermally grown layers exhibit lower values than those of chemically deposited films. In fact, the large range arises from differences induced during the deposition process, where defect or impurity concentrations can be modulated. For example, the hydroxyl (OH) radical has been attributed as the primary source of TLS in $SiO_2$ glass [87],[88],[110] and in amorphous Si, where its concentration was directly corelated to tunneling states [80],[86]. The molecular rotation of the OH group, due to its low coordination within its atomic environment, provides multiple minima in the energy landscape through which the TLS is formed. **Fig. 5a** depicts such a hypothetical arrangement, where the hydroxyl radical is bonded to a reconstructed Si (001) surface. Also contained in this figure is a hydrogen atom that can passivate the adjacent dangling bond, which itself can represent a TLS [79],[92]. Similarly, silicon nitride films also exhibit a range in loss tangents that depend on the deposition method. Measurements conducted using Al-based lumped element resonators revealed that films with a higher concentration of N in chemical vapor deposited (CVD) silicon nitride films exhibited greater loss tangents [106]. As shown in **Fig. 5b**, the bonding arrangement of the $NH_2$ complex to Si acts in a manner analogous to OH [111]-[113]. In addition, TLS dipole moments, as inferred from lumped element resonator measurements in $SiN_x$ films were comparable to those found in bulk silicon oxide [114], where dielectric loss is predicted to scale with the square of the dipole moment [88]. OH is also suspected to be a primary contributor to TLS in alumina [96]. Through density function theory (DFT) calculations, hydrogen impurities located on interstitial positions in α-Al₂O₃ have also been suggested as a TLS as they tunnel between the nearest and next nearest oxygen atoms [95].

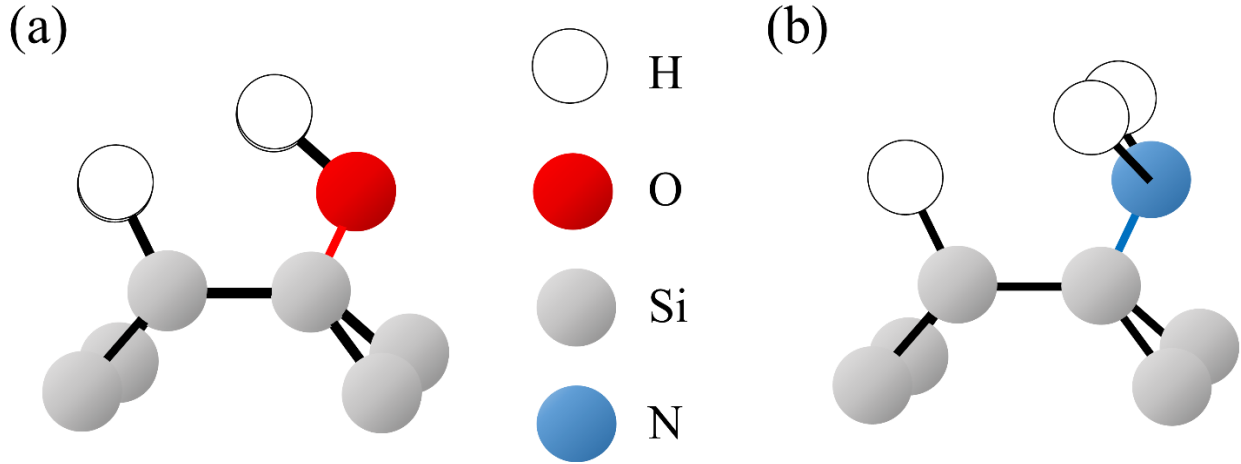

FIG. 5: Hypothetical schematic of (a) a hydroxyl group on a reconstructed Si (001) surface with the resulting dangling bond passivated by a hydrogen atom, and (b) nitrogen radical and hydrogen atom on a similar reconstructed Si (001) surface.

Thin films of materials exhibit larger loss tangents than their single-crystal counterparts, as can be seen in the values from alumina films and a sapphire substrate in **Fig. 4a** and **4b**, respectively, due to differences in defectivity and atomic coordination. The measurement of dielectric loss in bulk crystals often utilize superconducting cavities, in which whispering galler mode (WGM) resonances are excited and compared to EM simulations of the corresponding field distributions [107]. Using a combination of WGM resonance and electron spin resonance techniques, dielectric loss tangents of $2 \times 10^{-8}$ have been reported in high-purity, heat exchanger method (HEM) sapphire substrates at low power and 27 mK, with a corresponding magnetic loss tangent on the order of $10^{-8}$. This approach is extremely sensitive to paramagnetic impurities, such as Fe in the range of 100 ppb in addition to Cr and V [115]. Similar measurements on Si substrates confirm the improvement in loss tangent of isotopically purified $^{28}$Si substrates over float-zone, high-resistivity wafers, though neither exhibited values below 1.4 x 10$^{-6}$ at single photon powers and 20 mK [116]. Quality factor measurements based on superconducting resonators and qubits can also provide an upper bound on the underlying substrate's loss tangent. The procedure for calculating the appropriate amount of electromagnetic energy located in the substrate will be described in Section 3 but interestingly the dielectric loss extracted from this approach, as shown in **Fig. 4b**, is less than that reported in the previously mentioned WGM measurements.

### 2.3 Extrinsic effects

The results presented in the previous section on the effects of impurities implies that processing associated with superconducting resonator and qubit fabrication can have a deleterious impact on the dielectric loss within the substrates. In fact, a significant decrease in the quality factors of Nb CPW's manufactured on Si substrates was observed with decreasing substrate resistivity, presumably because of increased dopant concentrations [117]. **Fig. 4b** also shows the increase in loss tangent by several orders of magnitude that occurs when damaged top surfaces are produced by argon ion milling of sapphire [51],[104],[118] and silicon substrates [66]. However, avoiding ion milling prior to metal deposition can also result in reduced Q values of CPW resonators due to incomplete removal of underlying lossy dielectrics [104]. Such an example is illustrated in **Fig. 6**, where an oxygen ash 'descum' treatment reduces organic residue below the center traces of Al metallization CPW's fabricated using lift-off evaporation, leading to greater internal quality factors, $Q_i$, at low photon number. Even the use of freestanding, silicon-on-insulator

(SOI) layers as the dielectric layer in lumped element resonators [119] or in shunting capacitors for phase qubits [120],[121] suggest loss tangents (greater than $5 \times 10^{-6}$) significantly higher than those displayed in **Fig. 4b** for pristine single crystals. However, the deposition of epitaxial Si and SiGe layers on Si substrates prior to transmon qubit formation did not impact their performance [122]. Such findings underscore the delicate balance between preparing pristine substrate surfaces prior to as well as after qubit and resonator processing without inducing additional damage.

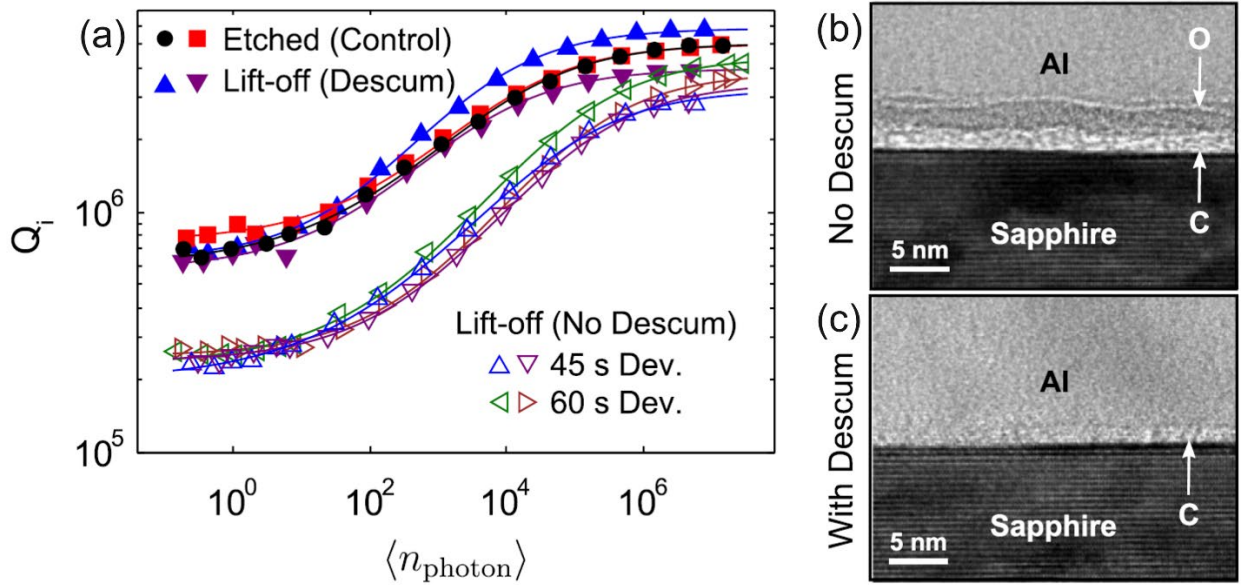

FIG. 6: (a) Comparison of CPW resonator internal quality factors, $Q_i$, as a function of photon number, illustrating the improvement induced by an oxygen ash descum treatment prior to center trace deposition which reduces remnant organic residue caused by lift-off evaporation. (b) Cross-sectional transmission electron microscope (XTEM) imaging of the center trace metallization / substrate interface revealing contamination without and (c) with the descum process. Reprinted courtesy Quintana et al. [104].

Methods of cleaning substrate surfaces include chemical routes (etching, passivation) and physical approaches (ion milling, heat treatments) as developed for conventional microelectronic fabrication. Prior to the deposition of metallization, these methods can involve solvent cleans such as acetone, followed by isopropanol, [51],[104] or methanol [123] and water immersion to remove incoming contaminants. For Si substrates, an RCA cleaning step [64],[124] is typically performed. High-temperature annealing of sapphire substrates both helps to remove organic residue and reconstruct the top surface, independent of the method of the overlying metallic deposition [125]. Heating of a Si substrate in the proper ambient also reduces its oxidation, leading to improved quality factors in Al CPW resonators, particularly in combination with chemical etching prior to annealing [126]. However, excessive thermal budgets (950 $^0$C) were shown to induce significant roughness of the Si surface and a reduction in CPW resonator Q values. In addition, silicon wafers can be treated with dilute or buffered HF solutions to remove the native oxide layer prior to metallization [63],[93],[108],[127],[128]. Passivation of Si substrate surfaces using hexamethyldisilazane (HDMS) creates a hydrophobic surface, so that its application immediately prior to metallization deposition may mitigate TLS formation at the substrate / metal interface, leading to higher quality factors observed in CPW resonators [63] and qubits [65]. Depending on the choice of metallization, HF etching can be applied after resonator formation, leading to a decrease in dielectric loss by reducing oxide contamination at the free Si substrate surface [129]. As will be described in greater detail in Section 4, the steps associated with specific deposition processes may also generate unwanted dielectric materials in the vicinity of these features.

## 2.4 Spectroscopic Characterization

Sections 2.2 and 2.3 highlighted the aggregate response of TLS to dielectric loss and its impact on superconducting qubits and resonators. Another method that attempts to identify TLS involves qubit spectroscopy, in which the resonant frequency of select qubits is swept across a range of interest by applying flux. Strong coupling between the energies of the qubit and TLS within or near the Josephson junction results in an avoided level crossing when their frequencies overlap [130]. In this manner, qubit performance can exhibit a discrete response in frequency to individual TLS [131]-[133] depending on their density [134]. Decoherence in phase qubits induced by TLS was recognized with this technique, where fewer avoided level crossings were found in small Josephson junctions compared to larger ones [86]. A similar approach can also be applied to lumped element resonators, where transmission across capacitors is modulated by TLS residing within the dielectric insulation under an applied voltage bias [114].

Recent extensions of this technique have revealed the sensitivity of qubit relaxation rate to the application of voltage bias or substrate deformation. By placing electrodes above and below the qubit chip, one can apply a DC voltage which is capable of modulating the resonance frequencies of a fraction of the TLS [135]. Electric field simulations suggest that the location of such TLS would need to reside within approximately 100 nm of the qubit electrodes [134],[136],[137] or within the tunnel barrier of the Josepshon junction itself. Also interesting is that inducing strain within the underlying substrate, through the use of a piezo actuator [138], shifts a greater fraction of TLS than does DC bias in these same qubits, as shown in **Fig. 7** [135]. The difference between these two populations has been ascribed to TLS within the Josephson junctions, whose large electric fields should dominate over those imposed by an external DC bias.

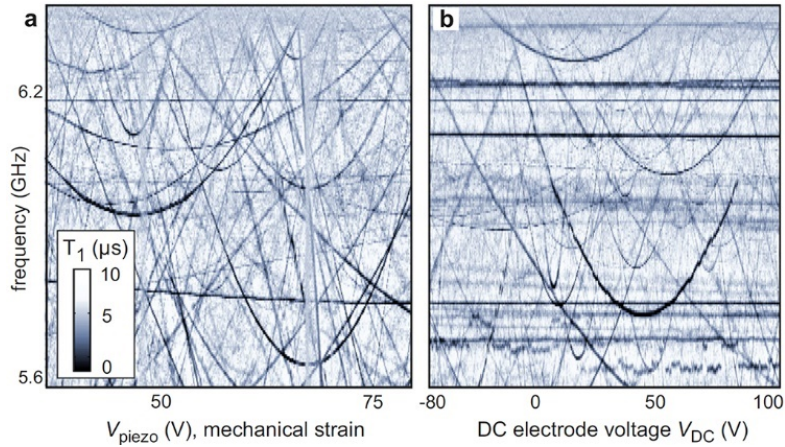

FIG. 7: Plots of TLS resonances which impact the relaxation times of flux-tunable, transmon qubits due to (a) substrate strain or (b) DC bias applied to electrodes above and below the substrate. Courtesy Lisenfeld et al. [135] licensed under a Creative Commons Attribution license.

**2.5 Summary**

Loss due to dielectric materials, caused by the interaction of TLS with the ambient electromagnetic environment, plays a major role in dictating the performance of superconducting qubits and resonators at low temperatures. Their ubiquitous nature makes it difficult to uniquely identify which species may be primarily responsible, though amorphous microstructures generally exhibit greater loss than their corresponding crystalline counterparts. Analytical characterization coupled with quality factor measurements suggest specific radicals (OH, $NH_2$), dangling bonds and carbonaceous residue as likely candidates. Advances in TLS spectroscopy, where coupling between TLS and qubits can be modulated through electric and mechanical manipulation, have helped to provide a means to localize the sources of such loss.

## 3. Participation

A key part of determining the susceptibility of qubit designs to dielectric loss is the calculation of the electric field energy that propagates through the neighboring dielectric materials. Treating the dielectric response of TLS using a continuum framework allows one to integrate the electric field energy, U, over select regions of the qubit and resonator designs which, when combined with the appropriate loss tangent, provides an evaluation of the impact of dielectric loss on the quality factors of these devices. Let us start by integrating the electric field energy density within a prescribed volume, $V_i$:

$$U_i = \frac{1}{2} \int_{V_i} \vec{E} \cdot \varepsilon_i \vec{E} \, dV \tag{4}$$

where $\varepsilon_i$ refers to the relative dielectric constant of the material. The participation of a particular region, P, also referred to as the filling factor [139], is simply the relative fraction of its energy, $U_i$, to the total electric field energy of the system, $U_{tot}$.

### 3.1 Key interfaces

Based on the previous section, the key portions of the design to be evaluated include the substrate-to-metal (SM) interface between the substrate and overlying metallization, the substrate-to-air (SA) interface corresponding to the free surface of the substrate and the metal-to-air (MA) interface along the exposed surfaces of the metallization. **Fig. 8** illustrates these various sections for both coplanar waveguide (CPW) and coplanar capacitor (CPC) geometries.

Note that the assessment of the participation in these regions could correspond to contamination layers or to defectivity induced in the underlying substrate due to metallization deposition or etching, under the assumption that the depth, $\delta$, of such regions is much less than the lengthscales associated with the coplanar designs, such as the CPW centerline conductor width (2a) in **Fig. 8a** or the CPC gap (2a) in **Fig. 8b**.

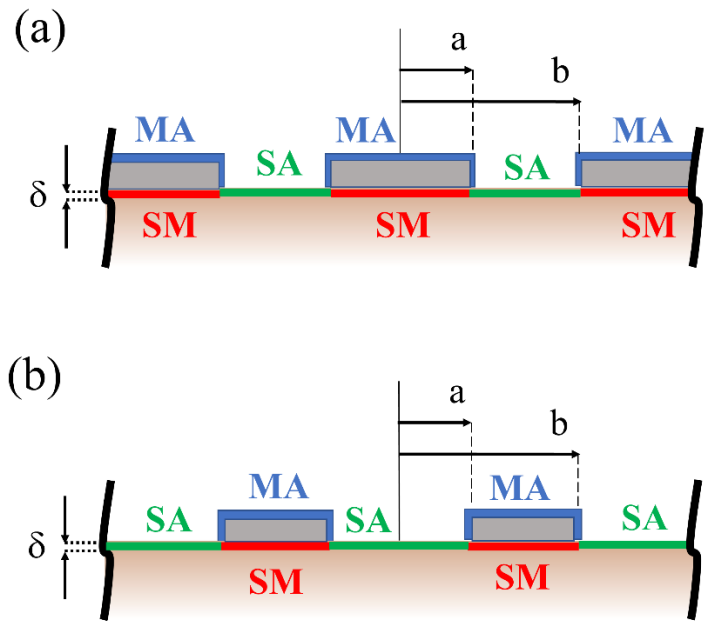

FIG. 8: Cross-sectional schematic of (a) CPW resonator and (b) qubit coplanar capacitor (CPC) designs illustrating key interfaces on the metallization (grey) and substrate (orange): substrate-to-metal (SM), substrate-to-air (SA) and metal-to-air (MA).

## 3.2 Substrate loss

The calculation of electric field energy within the substrate is the most straightforward of all the quantities, as it is derived in the same way as the capacitance, C, between superconducting elements ($U_C \sim CV^2/2$). Through conformal mapping, C.P. Wen [140] created a transformation from a parallel-plate cross-section to a CPW geometry to calculate its transmission characteristics, and an analogous transformation can be applied to coplanar capacitor designs [141],[142]. In the limit of a semi-infinite substrate, its participation can be approximated by the ratio of its relative dielectric constant: $p_{SUB} \sim \varepsilon_{SUB}/(\varepsilon_{SUB}+1)$ for untrenched geometries. For Si or sapphire substrates, $p_{SUB}$ is approximately 90%, much higher than any other portion of the superconducting qubit geometry. As discussed in the previous section, surfaces and interfaces of these substrates adjacent to the qubit metallization, and their modification due to processing, may be more relevant in a description of dielectric loss.

## 3.3 Modeling approaches

Approaches based on the finite element method (FEM) represent the most popular method to calculate participation [64],[108].[123],[143], as such models are extremely versatile with respect to the geometries they can analyze. However, the great disparity in lengthscale between the interfacial regions, which may be nanometers in thickness, and the overall dimensions of the design that can span hundreds of microns, complicates the process of domain discretization. Though surface participation is considered to describe the effects of impurities, defects, TLS residing on select surfaces, the determination of electric field energy typically involves integrating the energy density over a specific volume. Because the contamination layer thicknesses are usually not known a priori, Eq. 4 can be approximated by a surface integral multiplied by an arbitrary thickness, $\delta_i$ [108],[123],[144]. However, such surface participation, which may not vary linearly with layer thickness, also requires integration over singular electric field distributions that diverge at the metallization edges [142],[143],[145]. **Fig. 9** provides depictions of the calculated electric field energy at the substrate surface for three different qubit designs, where an increase in the gap size (2a in **Fig. 8b**) between shunting capacitor paddles or increasing paddle dimensions (b-a in **Fig. 8b**) reduces the electric field density, consistent with experimental evidence collected from resonators [118] and qubits [134].

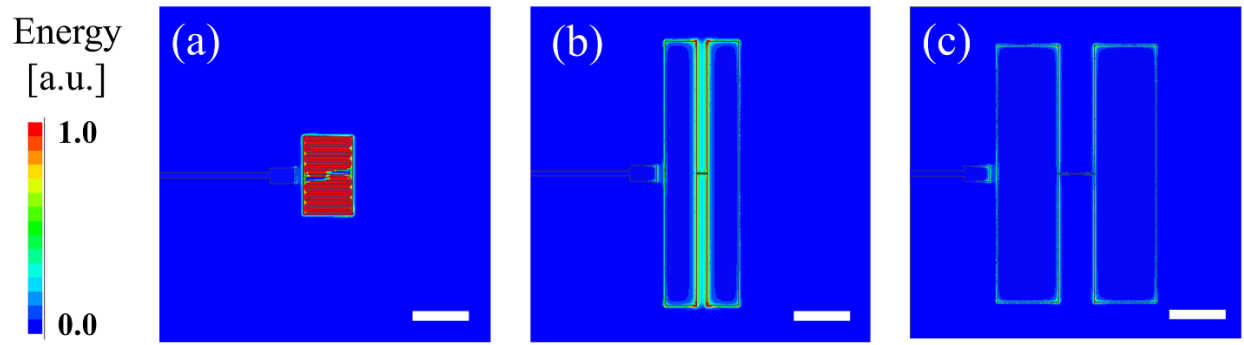

FIG. 9: Top-down view of the simulated, electric field energy distributions under qubit metallization possessing (a) interdigitated capacitor design (5 micron gap), and rectangular paddle designs with a (b) 20 micron gap and (c) 70 micron gap. The white scale bar corresponds to 100 microns.

Instead of FEM approaches, analytical modeling allows one to treat the geometry of interest in a holistic manner without the need to match truncated solutions from separate portions of the design [123] or power law approximations of the electric field intensities [64],[146]. Conformal mapping, a technique employed to estimate the capacitance of coplanar waveguides [140],[147], can also be successfully applied to the study of participation in designs that can be viewed as largely two-dimensional in nature. Taking a cross-section through a transmon qubit design that intersects the shunting capacitors (but not the leads to the Josephson junction) results in a profile similar to that of **Fig. 8b**. A quasi-static approximation of the electric potential, $\varphi$, which exhibits an antisymmetric profile with respect to the centerline, can be employed.

Although the corresponding electric fields possess singular forms as one approaches the metallization edges, through the use of Green's first identity [148], the volume integral represented in Eq. 4 can be transformed into a surface integral that converges when the contamination layer thickness, δ, is finite [142]. **Fig. 10** depicts the calculated participation at the SM, SA and MA interfaces for a coplanar capacitor design, revealing a slightly nonlinear dependence on δ. It is assumed that the relative dielectric constant of the contamination layer, $\varepsilon_{SA} = 5.0$ and that of the substrate, $\varepsilon_{Si} = 11.45$.

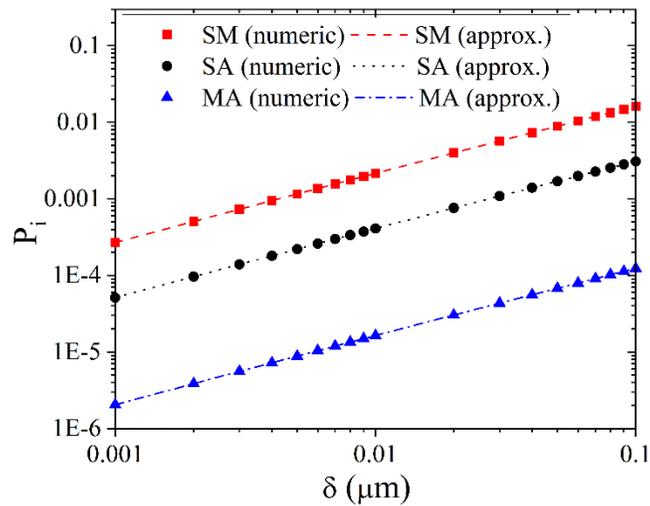

FIG. 10: Plot of simulated SM, SA and MA surface participation, $P_i$, of a qubit CPC design as a function of contamination layer thickness (δ) for paddle designs with a gap (2a) of 20 microns and widths of 60 microns. The relative dielectric constant of the contamination layer thickness is assumed to be 5.0. The lines correspond to the analytical approximation contained in Eq.'s 5 to 7. Courtesy Murray et al. [142].

A simple approximation based on the form of the electric fields in coplanar designs, under the assumption of metallization with zero thickness [142]:

$$P_{SA} \sim \frac{\varepsilon_{SA}}{(\varepsilon_{sub}+1)} \frac{1}{2(1-k)K'(k)K(k)} \cdot \left(\frac{\delta}{a}\right) \left\{ \ln\left[4\left(\frac{1-k}{1+k}\right)\right] - \frac{k\ln(k)}{(1+k)} + 1 - \ln\left(\frac{\delta}{a}\right) \right\} \quad (5)$$

is also plotted in **Fig. 10**, matching the values calculated using the surface integral method, where K and K' refer to the complete elliptic integral of the first kind and its complement, respectively, and k = a / b. From **Fig. 8**, it is apparent that CPW designs, where the metallization gap exists between the centerline conductor and groundplane, possess a complementary geometry to coplanar capacitor designs, where the location of the paddle metallization corresponds to the CPW gap. Although the electrostatics differ between these two geometries (a finite potential is assumed on the centerline conductor and zero on the left and right groundplanes of the CPW while an antisymmetric potential is used for the CPC), Eq. 5 can be used for both CPW and CPC cases, where 2a refers to the centerline conductor width or capacitor gap, respectively, and likewise 2b the distance between the edges of the groundplane or outer edges of the capacitor paddles, respectively. Eq. 5 also illustrates the nonlinear dependence of participation on the key length scales associated with the design (gap size, contamination layer thickness), demonstrating that a simple, power-law approximation does not hold [149].

Participation at the other interfaces can be approximated by considering the boundary conditions associated with EM radiation at dielectric interfaces. In the limit of thin contamination layers, the electric fields in regions directly below and above metallization will be oriented normal to the metallization / dielectric interface, while that in the gap region runs parallel to the SA interface. If we consider a contamination layer present at any of these interfaces, then the magnitude of the electric field will be modulated by the relative dielectric constants of this layer and the adjacent dielectric, along with the orientation of the fields across their interface [108],[145]. For contamination layers with identical thicknesses [142]:

$$P_{SM} \sim \frac{\varepsilon_{sub}^2}{\varepsilon_{SM}\varepsilon_{SA}} P_{SA} \qquad (6)$$

$$P_{MA} \sim \frac{\varepsilon_{SM}}{\varepsilon_{MA}(\varepsilon_{sub}^2)} P_{SM} \qquad (7)$$

Thus, for coplanar designs, these participation values (SM, SA, MA) can be treated as linearly dependent [142], which makes it difficult to uniquely identify which surface is responsible for dielectric loss when a comparison is attempted between untrenched resonators or qubits with different design dimensions. While SA participation is predicted to scale with $\varepsilon_{SA}$ (Eq. 5), Eq.'s 6 and 7 demonstrate that SM and MA participation decrease with increasing dielectric constant of their respective contamination layers. The difference is simply a consequence of the boundary conditions: electric fields radiating from the superconducting metallization traverse normal to the SM and MA interfaces while those associated with SA primarily travel parallel to the interface. Because the substrate relative dielectric constants are greater than 10 for sapphire and silicon, we can estimate the relative importance of the various interfaces: SM and SA participation values are much larger than MA in most cases (**Fig. 10**). Note that, in the case of metallization with finite thickness, MA participation is expected to be approximately twice as large as that predicted in Eq. 7, due to a doubling of the number of metallization corners, at which the electric fields are maximized, from 2 for each infinitely thin paddle to 4. However, some FEM-based modeling approaches report even larger MA participation values [58] which can result when the MA contamination layer is placed in direct contact with the underlying substrate. This creation of 'SM-like' interfaces augments the participation in such localized regions, again as prescribed by the boundary conditions associated with the MA contamination / substrate interface, which can dominate the total, calculated MA participation [122].

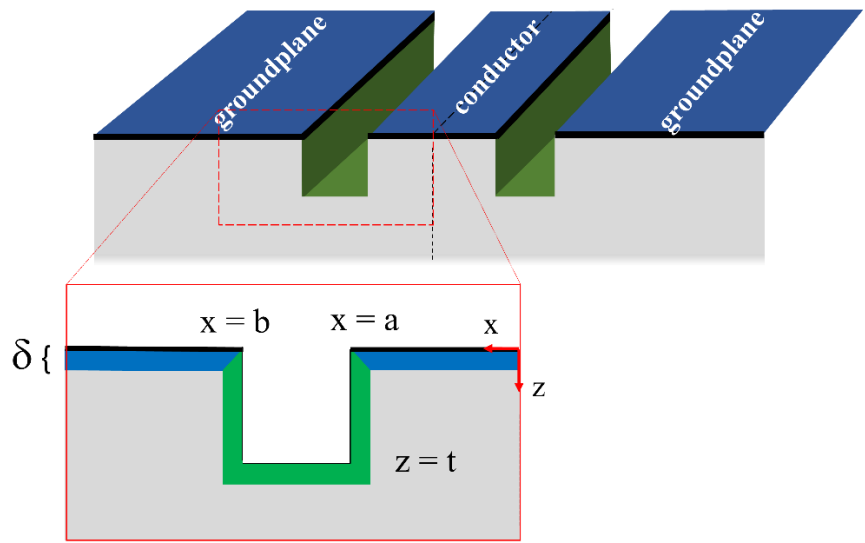

FIG. 11: Cross-sectional schematic of a CPW resonator geometry upon substrates with trenches of depth, t, in the gap between the groundplane and centerline conductor metallization.

**3.4 Effects of substrate geometry**

One method that can reduce participation among all of the various interfaces in resonator and qubit designs involves removing portions of the dielectric substrate between the overlying metallization [63],[91],[150]-[152]. As illustrated in **Fig. 11**, the creation of recesses or trenches within substrates, which reduces its effective dielectric constant, is more facile in certain materials, such as silicon, due to its reactivity with several chemical species whereas others like sapphire are more resilient to etching.

FEM approaches are again commonly used to simulate the impact of trenching on participation [64],[143],[153]. Singularities in the electric fields in close proximity to the metallization edges still complicate such analysis and are either disregarded [153] or treated by power law approximations [143]. These regions must be included for an accurate accounting of the total electric field energy. Utilizing conformal mapping, we can generate an analytical solution by effectively flattening the trenched substrate geometry to one with transformed conductor widths and gaps but a non-uniform contamination layer thickness [154]. Results from such a model are shown in **Fig.'s 12a** and **12b**, at trench depths much smaller than those which can be easily simulated using FEM approaches. Although it is clear that the ratio of SM to SA participation now varies as a function of trench depth, the ratio of SM participation to that along the SA trench bottom only is still dictated by Eq. 6 for intermediate trench depths, consistent with untrenched geometries. Participation values calculated for trenched substrates indicate that their reduction is anticipated to saturate at trench depths greater than 4 to 5 times the coplanar capacitor gap **(Fig. 8b)** or centerline conductor width in coplanar waveguide designs **(Fig. 8a).** In the case of the latter geometry, an analogous value of 10 times the gap in CPW geometries was observed with gaps half of the centerline conductor width [64]. Below these depths, the electric field energy emanating from the metallization does not appreciably interact with the portion of the substrate below the trench bottom.

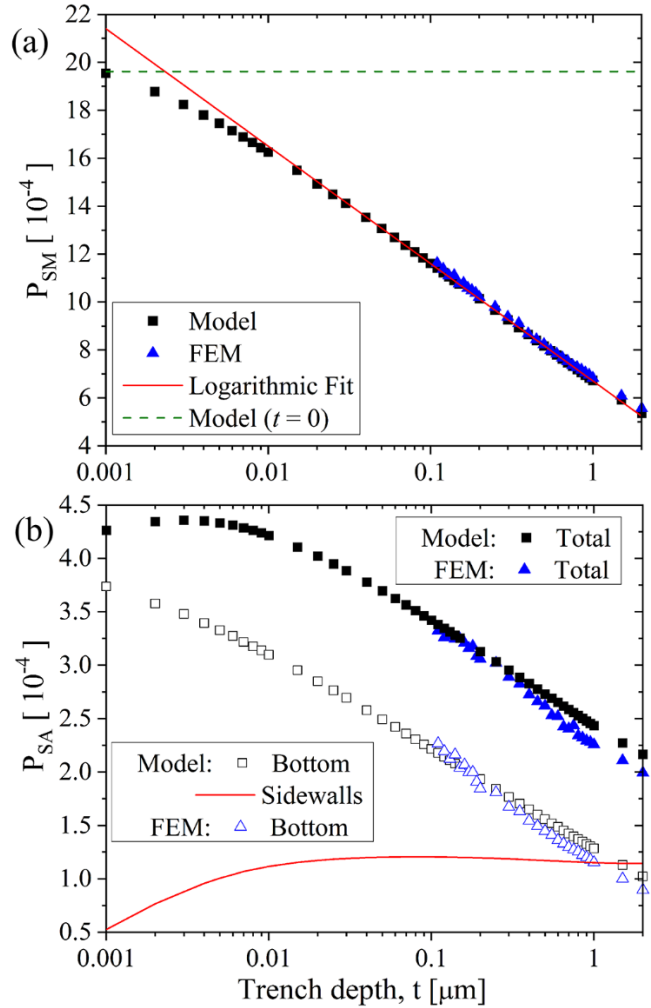

FIG. 12: Plot of simulated surface participation values as a function of trench depth, t, of a CPW design with a centerline conductor width (2a) of 10 microns and a gap of 6 microns. (a) SM surface participation that exhibits a logarithmic dependence (solid line) for depths greater than 20 nm and saturates at the value (dashed line) governed by Eq.'s 5 and 6. (b) SA surface participation whose fraction from the trench sidewalls (solid line) saturates at trench depths greater than approximately 50 nm. Courtesy Murray [154].

Although this approach assumes that the trench sidewalls are orthogonal to the trench bottom, simulations have been conducted on the effect of the angle at the trench corners [155]. For a given trench depth, surface participation is predicted to decrease further as the angle at the trench bottoms becomes more acute, reducing the effective dielectric constant of the substrate by incorporating a greater portion of vacuum underneath the metallization. In a complementary manner, a more obtuse trench corner angle will increase surface participation, ultimately approaching the values corresponding to an untrenched geometry.

### 3.5 Comparison to experimental measurements

Several efforts have been undertaken to extract loss tangents on key interfaces of qubit designs based on the participation factors as determined using the procedures outlined in the previous subsections. The quality factor is impacted by the sum of parallel dielectric loss channels associated with each of the interfaces:

$$\frac{1}{Q} = \sum_i P_i \tan(\delta_i) + \frac{1}{Q_0} \tag{8}$$

where $Q_0$ refers to a participation-independent term. Eq. 8 allows one to construct an algebraic expression relating the SM, SA and MA loss tangents as weighted by their corresponding participation factors, $P_i$ [123]. Alternatively, upper bounds for these loss tangents can be calculated by assuming that a particular loss channel dominates the observed qubit behavior. In this manner, loss tangents of $2.6 \times 10^{-3}$, $2.2 \times 10^{-3}$ or $2.6 \times 10^{-2}$ could be ascribed to the SM, SA or MA interfaces respectively for three-dimensional, evaporated Al qubits on sapphire under the assumption of 2 nm thick contamination layers [123]. For qubits consisting of Nb capacitor paddles and Al / AlO$_x$ / Al junctions, a comparable analysis provides an upper bound of approximately $2.0 \times 10^{-3}$ [108] and $2.6 \times 10^{-3}$ [122] at the SA interface, consistent with values associated with oxide contamination.  A lower bound on the effective loss in CPW resonators was also mapped by surveying results collected from a variety of metallization choices that scales nonlinearly with the gap width [53].

In regard to substrate trenching, recent investigations of CPW resonators have employed a variety of etching profiles within the underlying Si substrate in order to affect the linear dependence of the surface participation parameters [129],[144].  This approach, in concert with Monte Carlo simulations to refine the matrix inversion procedure, produces a SA loss tangent of $1.7(4) \times 10^{-3}$, again in accord with that expected for a silicon oxide contamination layer.  It should be noted that the increase in experimental quality factors measured in trenched CPW's has been observed to saturate at more modest trench depths [153] than 4 to 5 times the centerline conductor width.  Since participation is predicted to decrease monotonically [144] at much deeper trenches, as shown in **Fig.'s 12a** and **12b**, it suggests that either dielectric loss no longer becomes the dominant mechanism for CPW performance, or that loss tangents associated with the sidewall surfaces could be increasing, perhaps due to damage induced by prolonged etching [65].  However, removal of the underlying Si substrate through a Bosch etch process to create suspended qubits, generated the maximum Q values, approximately $2 \times 10^6$, associated with the study [72].

## 3.6  Summary

Continuum representations of dielectric loss have been employed to assess the susceptibility of superconducting qubit and resonator structures.  Participation, the relative fraction of electric field energy flowing through key surfaces and interfaces within qubit architectures, is combined with experimentally determined loss tangents to quantify dielectric loss. Analytical approximations, which capture the effects of electric field singularities at metallization edges better than finite element method formulations, allow one to identify design criteria that can help to reduce participation, such as increasing the distance between capacitor metallization and recessing dielectric material within these gaps. However, experimental verification of these approaches points to a possible saturation of their benefit to qubit and resonator performance, suggesting that other loss mechanisms may start to dominate.

## 4. Metallization

A variety of metallic films, consisting of elemental or alloy superconductors, has been employed in the fabrication of superconducting qubits. Either one or a combination of materials can comprise the metallization associated with resonators, inductors, capacitors and groundplanes. These choices exhibit a wide range in intrinsic properties, such as superconducting transition temperature, London penetration depth and normal state resistivity. However, the level of impurities and disorder present in such films can play just as important a role in dictating their overall behavior. Therefore, how these materials are deposited and subsequently processed may impact the quality factors associated with their application in qubit architectures.

### 4.1 Composition

Aluminum represents the most prolific example of metallization incorporated in quantum computing applications [130],[156], including one of the first example of Cooper pair boxes created with shadow evaporated Al on a silicon nitride dielectric layer located above a gold groundplane [44]. Its use in the fabrication of Josephson junctions spans an even longer history due to the formation of a robust, passivating oxide with a thickness appropriate for Cooper pair tunneling [157],[158]. As shown in **Fig. 3a** and **3b**, several methods of depositing Al have been employed in an effort to optimize qubit and resonator performance. Each of these techniques have their strengths and disadvantages; for example, molecular beam epitaxy (MBE) can produce atomically flat films but at a much slower rate that those associated with physical or chemical vapor deposition. Although evaporated Al may appear to outperform resonators constructed using other methods, this comparison could instead reflect ancillary effects associated with their processing, as discussed in Section 4.3.

As refinements in qubit designs produced longer coherence times [5],[31], alternative metals were introduced to investigate if the intrinsic properties of aluminum were limiting qubit performance. For example, a superconducting transition temperature, $T_c$, of approximately 1.2 K make aluminum more susceptible to effects related to its lower superconducting gap energy (see Section 6). While elemental Group 4 metals (Hg, Pb, Sn) were explored in early adaptations of superconducting electronics [159], their low melting temperatures make these choices more sensitive to conventional microelectronic fabrication processes. Niobium has been incorporated in qubits [160] and other superconducting applications [161] due to its $T_c$ of approximately 9.5 K and critical magnetic field of 0.2 Tesla, the highest of any elemental superconductor [162]. Interestingly, the superior nature of oxidized Al as a tunnel barrier (as discussed in Section 5) motivated its inclusion in qubit builds for a variety of alternate superconductors as shunting capacitor, resonator and groundplane metallization, such as Nb [89],[163]-[165].

Recent investigations have involved the use of tantalum, another Group 5B element possessing a relatively high $T_c$ (4.5 K), for qubit and resonator metallization. As seen in **Fig. 3b**, the corresponding quality factors of Ta qubits fabricated on sapphire substrates are among the highest reported in literature [69]. Its insulating, native oxide layer ($Ta_2O_5$) is hypothesized to generate lower loss [166] than those associated with the several possible oxidation states present on Nb films [167]. One challenge is that the metastable, β-Ta phase, which generally forms by sputtering, possesses a much lower $T_c$ than that of the equilibrium, alpha phase so that stabilizing the body-centered-cubic (BCC) phase requires an underlayer [168],[169] or sputter deposition at elevated temperatures [170]. The high-deposition temperature was also suggested as important to generate a microstructure within the Ta films containing fewer diffusion paths for oxygen to ingress [69].

Nitrides formed with transition metals from Group 4B and 5B have also been investigated, where the cubic (rock salt) structure corresponds to a superconducting phase [171]. The most ubiquitous of these found in resonators and qubits is titanium nitride, which is predominantly created through sputtering [89],[93], though CVD [172],[173], pulsed laser deposition [174] and atomic layer deposition (ALD) [175],[176] have also been employed. Niobium nitride (NbN) and niobium titanium nitride (NbTiN) have also been utilized [177], which possess $T_c$ values as high as 16 K [178]. However, the atomic spacing may represent a more important property, as it affords the possibility of epitaxial alignment with select substrates, discussed in Section 4.2. Thin films of such materials can possess long London penetration depths [179],[180] and correspondingly large kinetic inductances, making them highly desirable in photon detector applications [181],[182] as well as quantum computing architectures that require high inductances [42],[54],[183].

A plethora of other candidates exist, such as rhenium [68],[184], vanadium [185], hafnium [186], indium [187], molybdenium-rhenium alloys [188], compounds that exhibit the A15 crystal structure [189] and even heavily-doped silicon [190], that have been demonstrated or proposed for use in quantum computing. While some of these materials also possess superconducting transition temperatures above 10 K, the primary motivation of their incorporation is not necessarily associated with $T_c$, since the temperature necessary to reach the qubit ground state ($k_B T \ll hf$) is substantially lower. According to **Fig. 3b**, the fact that Ta qubits exhibit larger Q values than Nb suggests that the absolute value of $T_c$ does not dictate quality factor. As discussed in the next sections, other properties of these materials may be more relevant in defining device performance.

**4.2 Texture**

When considering the effects of crystal orientation, or texture, within superconducting metallization, it is important to distinguish between two aspects: the assemblage of grain orientations in a polycrystalline microstructure and the usage of epitaxy as a template in the deposition of subsequent layers. In the former case, conventional deposition of metals and alloys often produces polycrystalline films. In the absence of an underlying epitaxial template, close-packed planes will lie parallel to high-surface energy deposition surfaces in an effort to minimize surface energy: metals which possess a face-centered-cubic (FCC) unit cell prefer (111) texture while those with a body-centered cubic arrangement exhibit (110) texture. Depending on the growth conditions (temperature, pressure, surface treatment), Al (110) can form on Si (001) [191],[192] while Al (111) can be formed on Si (111) [193]-[195]. Another consideration is the heterogeneity in oxide thickness on the particular crystal surfaces within the underlying metallic film [196] which, as discussed in Section 4.3, may produce differences in dielectric loss.

Measurements performed on coplanar waveguide resonators composed of TiN metallization have ascribed differences in low power, internal quality factors to its texture, where films composed of predominantly (200) orientation exhibited at least a factor of two improvement over those possessing (111) texture [93]. The challenge is delineating between causal relationships of texture on performance and correlational observations related to microstructural effects. The superconducting gap of most simple, metallic superconductors is isotropic, so that their intrinsic properties are anticipated to be insensitive to texture. Because the reported resistivities of the metallization differ by nearly a factor of four between growth of TiN (200) at elevated temperatures (500 $^0$C) upon a nitridized silicon surface and that at room temperature on Si substrates, it suggests that impurities, such as oxygen, or a difference in film density or in grain size may also be present. Magnetron sputter deposited TiN films at 500 $^0$C revealed a decrease in oxygen content and a similar change in texture from (111) to (200) with increasing DC bias [197] and comparable single photon quality factors (~3 x $10^5$) in (200)-oriented TiN resonators [93]. In fact, measurements conducted on TiN CPW's fabricated using increased $N_2$ pressure exhibited larger, low power quality factors ($10^6$) under conditions that promoted (111) texture and higher contaminant concentration [128].

Generating heteroepitaxial arrangements of superconducting metallization on select substrates provides a path to minimizing microstructural defects, such as grain boundaries, present in polycrystalline films. For example, Re (0002) films are closely matched to c-plane sapphire substrates, but with a $30^0$ in-plane offset [103]. CPW resonators composed of Re [198] and Al [199] have been fabricated on sapphire through the use of elevated temperature depositions. Al (111) films have been grown on both c-plane sapphire and Si (111) substrates [199], though the large mismatch in equilibrium lattice parameter can induce misfit dislocations and twin boundaries [200]. In fact, specific reconstructions of the top Si surface when exposed to high temperature prior to deposition is pivotal in mitigating defectivity in the overlying Al [195]. Plasma-assisted MBE [201] and ALD [176] deposition of TiN (111) on Si (111) have also produced CPW resonators with single-photon quality factors approaching $10^6$. It should be noted that epitaxial resonators do not exhibit superior quality factors to those shown in **Fig. 3a**, suggesting that polycrystalline microstructures within the superconducting metallization do not represent the dominant source of loss. However, the most promising application of these arrangements may be the creation of epitaxial tunnel junctions, as discussed in Section 5.

## 4.3 Processing effects

Analogous to the issues mentioned in Section 2.3 with respect to substrates, the intrinsic properties of superconducting metallization may be less important than the effects associated with fabrication of resonators or qubit structures in dictating their performance. For example, it has been demonstrated that lift-off deposition processes that incorporate organic templates can reduce the Q values measured in Al CPW resonators [66]. It is possible that organic residue remaining after the removal of the etch mask may simply be responsible [104]. In fact, exposing these structures to an oxygen plasma prior to metallic deposition can reduce such residue [202], and has been shown to improve Al resonators on sapphire substrates [66]. However, all surfaces must be considered when applying such techniques, as oxygen plasma treatment of TiN CPW's on Si substrates induced an increase in the MA loss tangent when the metallization was etched prior to its lithographic patterning [144]. A similar observation was found in Nb CPW resonators, where quality factors were observed to increase as the overlying Nb oxide thickness decreased through prolonged buffered oxide etching [57]. Although the regrowth of a thin (~ 1 nm) oxide layer took place, its loss tangent appeared to be much lower than that present after patterning, lending credence to the hypothesis that $NbO_x$ layers can impact resonator performance[58],[69].

When oxygen plasmas are applied to resonator structures after their formation, a significant increase in oxygen content was observed on the free Si surface, and a corresponding decrease of Q by a factor of 3 relative to those that underwent an HF etch of the substrates [152]. However, a similar degradation in performance was not found from freestanding Al qubits, in which Bosch etching generated over 60 micron deep trenches in the underlying Si substrate, possibly because the participation of the SA surfaces in such a structure was sufficiently small to mitigate loss due to oxide formation [72]. Nevertheless, approaches have been pursued to eliminate damage induced by residue or subtractive etching by employing an inorganic hardmask to evaporate Al features on sapphire substrates [203].

Additional complications can arise in the fabrication of superconducting nitride layers, particularly on silicon substrates. A large variation in nitrogen content within TiN films from the stoichiometric phase is possible, which can lead to a distribution in $T_c$. One method to mitigate this effect is to employ a multilayer film stack, where a secondary material is proximitized by the TiN which lowers the composite $T_c$ to a more uniform value [204]. However, the amount of nitrogen pressure during sputter deposition presents other consequences to resonator performance. For example, an increase in pressure during TiN deposition led to larger quality factors in resonators which contained a greater fraction of impurities and stronger (111) texture [128]. Because these films also exhibited less compressive stress with increasing nitrogen pressure, it is possible that a greater degree of atomic peening [205] can influence the underlying substrate surface. The role of the silicon nitride interlayer, which forms during initial exposure of the substrate to the deposition atmosphere, must also be considered. At low power, resonators experience dielectric loss due to buffer $SiN_x$ layers but the high-field performance is optimized [93]. However, similar to the discussion regarding texture (Section 2.2), these TiN film depositions employed high temperatures which reduced impurity and defectivity levels, as evidenced by the corresponding decrease in calculated London penetration depth values [93].

Given the high electric field intensities near the metallization edges, the question is often asked as to the impact of line edge roughness on device performance. From an analysis of power-dependent fluctuations in the dielectric loss of Al CPW's etched on sapphire substrates, an estimate of the TLS density of approximately $2/\mu m^3$ along the resonator edges was obtained [206]. Although Al-based, $\lambda/4$ CPW resonators with Q values greater than $10^6$ exhibited low defectivity along metallization edges during patterning, their performance did not appear to scale with defect density [199]. Measurements on sputter deposited, Nb resonators and qubits showed that aggressive $SF_6$ etching increased loss relative to those possessing tapered Si sidewalls, which could be explained by roughened Nb edges or possibly greater SA damage [65]. TiN CPW resonator loss on Si substrates was found to be greater when using chlorine-based reactive ion etching (RIE) as opposed to fluorine chemistries, where speculation as to the mechanism involved greater ion damage and the presence of boron in the chlorine etch gas [151]. Resonator etching using Ar ion milling produced fence-like features at the metallization edges, presumably due to redeposition of Si onto the photoresist sidewalls, and the largest loss [151]. One method to circumvent RIE damage is to employ wet etching techniques, which can lead to removal of fence-like features along the periphery of Ta metallization deposited on sapphire substrates [69]. Overhanging structures at the edges of Ar / $SF_6$ etched, Al-capped TiN resonators were also blamed for a reduction in Q values, suggesting that establishing optimal RIE process windows are essential in obtaining high quality factors [152].

The prevalence of combining superconducting metallization with aluminum, in order to form $AlO_x$-based Josephson junctions, places stringent requirements on the nature of their common interface. Since any passivation of the underlying surface can impede supercurrent flow, a galvanic connection is necessary to eliminate any parasitic junctions. In-situ Ar ion milling prior to the second metal deposition can be used to remove native oxide, leading to reduced loss in composite Al resonators [67] and Nb CPW resonators possessing Al segments in series with the centerline conductor [207],[208]. Because **Fig. 4b** demonstrates that such ion milling can damage the substrate and induce significant dielectric loss, 'bandage' connections were developed, where an additional conductive strip is placed on top of the metallization which received ion milling but not the adjacent substrate [66],[209],[210]. Galvanic connections are also crucial for metallic airbridges fabricated on groundplanes to mitigate the effects of slotline modes. Aluminum-based airbridges, formed by using a sacrificial silicon oxide layer subsequently removed through an HF vapor treatment, have been demonstrated in flux qubits [211]. However, degradation in resonator Q values appeared to scale with the number of airbridges [47],[212], suggesting that dielectric loss associated with their construction still remained.

The presence of strain within films has been proposed as a mechanism that impacts their performance in superconducting electronics applications [128],[213]. While greater disorder [177] and extreme pressure [214],[215] can affect the $T_c$ values of superconducting materials, it is difficult to decouple microstructural differences, which can be generated due to processing conditions (e.g: ambient pressure and voltage bias) [216], from differences directly attributable to residual strain in the films. For example, TiN CPW resonators fabricated using DC magnetron sputtering exhibited an increase in resistivity, oxygen, carbon content and low power quality factor with greater nitrogen pressure [128]. The authors ascribed this trend to the residual stress state in the TiN film, which transitioned from highly compressive to slightly tensile. However, it is also possible that films with the largest compressive stress, and low quality factors, may have experienced increased atomic peening of the Ar neutral atoms [205] that damaged the underlying Si surface during the initial stages of sputter deposition [217]. In contrast, measurements conducted on TiN CPW's fabricated on Si substrates demonstrated the opposite tendency that higher quality factors could be achieved with films exhibiting high compressive stress [197].

Although definitive evidence of strain within metallization impacting superconducting qubit performance may be lacking, deformation induced within the underlying dielectric substrate materials can alter the interaction between qubits and TLS [135],[218],[219]. Such strain, concentrated near the edges of the structures, may be substantial at cryogenic conditions due to differences in the coefficient of thermal expansion between the substrate and the overlying metallic features [220]. In this temperature regime, dislocation-based plasticity is expected to dominate deformation within many types of metallization [221],[222]. Resonant scattering of thermal phonons by dislocations has been reported in aluminum [223] and Nb [224] through lattice thermal conductivity measurements. However, strain fields induced by sessile dislocations are not suggested to be responsible for phonon scattering but rather a dynamic interaction must occur.

## 4.4 Inductive loss

The presence of a London penetration depth leads to portions within superconducting metallization that can absorb AC electromagnetic energy. The model of Mattis and Bardeen [225] describes such regions through the adoption of a complex surface impedance, $Z_s = R_s + iX_s$, where the resistance, $R_s$, is proportional to the dissipation [226]. Analogous to the dielectric loss terms contained in Eq. 8, inductive loss can be thought of as the product of the magnetic field energy participation within the penetration depth and a material dependent quantity, $R_s/X_s$ [53],[227], which is predicted to increase with frequency [226]. This mechanism is related to losses associated with quasiparticle generation which will be discussed in greater detail in Section 6, and is clearly apparent in three-dimensional, superconducting resonator cavities, where treatments to the internal surfaces can dramatically impact their observed quality factors

[62],[228]-[230]. Power can also be dissipated by the interaction of surface currents with cavity seams [231]-[233]. A phenomenological description of this process utilizes an effective admittance that is proportional to the inductive loss [231]. This admittance can be minimized by modifying the cavity geometry for a given resonator mode, so that surface currents flow parallel to rather than across the seams, or by improving seam quality, through, for example, incorporating indium bump bonds to indium groundplanes that coat micromachined Si cavities [233].

Paramagnetic surface impurities and defects can enhance such loss in resonators. Although their concentrations are typically low in single crystal substrates [115], oxygen vacancies on the oxidized Nb surfaces [57] and physisorbed hydrogen on NbN were correlated with reduced resonator quality factors, particularly at high photon number. These species were also linked to increased flux noise in superconducting resonators and SQUID's (see Section 7.2). In fact, the application of in-plane magnetic fields to NbTiN resonators generated sharp decreases in their internal quality factors at frequencies corresponding to $g = 2$ spin defects [234].

**4.5 Summary**

Choices abound in the selection of superconducting metallization for use in quantum computing, that span a large range of superconducting transition temperatures and a variety of deposition techniques. Despite these differences, recent examples of qubit and resonator fabrication have demonstrated a closer parity among their performance, suggesting that such intrinsic properties may currently play a secondary role as compared to processing effects (etching, dielectric contamination, etc.). While improvement in the microstructural quality of metallization has been achieved, it is not immediately clear whether these changes demonstrably influence the corresponding quality factors, and may require further mitigation of mechanisms such as dielectric loss before they become apparent.

## 5. Tunneling barriers

### 5.1 Composition

The creation of junctions between superconducting, metallic features have provided the realization of Josephson effects, launching applications that pervaded superconducting electronics approximately six decades ago. It represents one of the first manifestations of quantum effects on a macroscopic scale [235],[236] though Cooper pair tunneling has also been observed in atomic systems [237]. The primary motivation of incorporating Josephson junctions in superconducting qubits is that they provide a lossless, nonlinear inductance which, in combination with the overall capacitance, dictates qubit frequency. The most expedient method of generating Josephson junction barriers, oxidation of the superconducting leads, still remains the most popular. However, advances in lithographic patterning have allowed the formation of junctions using the direct deposition of dielectric materials, which also provides the ability to generate different microstructures such as epitaxial configurations.

As discussed in the previous sections, while numerous materials can comprise the shunting capacitors of superconducting qubits, the most ubiquitous choice of Josephson junction composition is aluminum oxide between aluminum leads. Although combinations involving the oxidation of Pb and Sn were some of the first in which Josephson effects were observed [238], those that incorporated a thin layer of oxidized Al within Nb on the bottom section of Nb cross stripes exhibited superior tunneling barrier quality than those using only Nb and Nb oxide [239]. Nb / Al / AlO$_x$ / Nb layers represent a common arrangement even for contemporary SQUID's, though increased subgap leakage and higher transparency was identified in such junctions, possibly due to pinholes in the barrier layer [240]. More recent investigations and simulations point to hybridization of the metallic Nb and Al$_2$O$_3$ conduction bands as being responsible for generating metal-induced gap states (MIGS) and reducing the potential barrier of the insulating barrier [241] when the two layers are in contact [242]. Increased subgap current was also observed with vanadium as the top electrode material for both Al$_2$O$_3$ [242] and MgO insulating layers [185].

Aluminum nitride (AlN) has been explored as a tunnel barrier for use with NbN metallization on Si substrates, where capacitively shunted flux qubits reported average quality factors of approximately 0.7 x 10$^6$ [243], 50 times greater than earlier attempts that incorporated higher loss MgO substrates [244]. Although piezoelectric effects in bulk and amorphous AlN should induce decoherence through phonon radiation [245], it is suggested that an epitaxial deposition process can stabilize the cubic phase to mitigate these effects. More exotic combinations include metal silicide layers (MoSi$_2$, TiSi$_2$, WSi$_2$) with Nb metallization as Josephson junctions [246] but, as these silicides are superconducting, they would not serve as insulating barriers at the temperatures necessary for superconducting qubit applications. Voltage-tunable transmon qubits employing van der Waals heterostructures, composed of graphene encapsulated by hexagonal boron nitride layers, were recently demonstrated [247]. Trilayer Josephson junctions fabricated using sputtered Si and Nb metallization have been fabricated [248] and incorporated into merged element transmon qubits [249], though the extracted short $T_1$ values (55 ns) were attributed to the amorphous nature of the Si layer.

**5.2 Formation**

The primary approach of creating aluminum oxide tunnel barriers for Josephson junctions involves the oxidation of exposed metallization prior to the deposition of a subsequent layer. Metal evaporation can be performed at two oblique angles through a template formed from organic layers using orthogonal in-plane directions ('Manhattan' style) [250],[251] or underneath a suspended ('Dolan' style) bridge at two opposing directions [252]. These deposition steps can be performed on top of existing metallic features, such as the metallization schemes contained in **Fig. 3b**, which comprise the other qubit components (shunting capacitors, resonators and groundplane). The potential impediment of 'parasitic' junctions [135] generated at the common interface of the two types of metallization can be addressed by in-situ Ar ion milling or metallic 'bandages' [66],[209],[210], as described in Section 4.3. In an effort to circumvent the two angle deposition process, two, sequential lithographic steps associated with each of the shadow evaporated Al layers have also been attempted, but also requires Ar ion milling in the junction region prior to junction formation [253],[254].

    The rate of oxide passivation on aluminum films depends on a complex relationship that involves temperature, oxygen pressure and oxidation time. Near room temperature, Cabrera-

Mott theory predicts a rapid, initial oxidation rate due to a self-generated electric field that reduces the potential barrier to ionic diffusion of chemisorbed oxygen which ceases at a limiting thickness when conduction electrons from the underlying metal can no longer tunnel through the barrier [255]. For a given superconducting energy gap within the leads, the Ambegaokar-Baratoff formula [256] predicts that the product of the normal-state tunneling resistance ($R_n$) and junction area will scale inversely with its critical current density, $J_c$. Experimental observations on Nb / Al / AlO$_x$ / Nb junctions confirm the trends of an exponential decrease of $J_c$ with the product of oxidation pressure and time that manifests itself in two regimes corresponding to low and high oxygen dose [257]. This behavior can also be modulated by other factors, such as electron beam bombardment [258] which is proposed as a mechanism to drive chemisorbed oxygen into vacancy sites within the aluminum oxide. An oxygen plasma ashing step prior to Al deposition was reported to reduce the variation in the corresponding Josephson junction properties, as well as a doubling of $J_c$ [251]. However, sub-stoichiometric AlO$_x$ ($x < 1.5$) Josephson junctions typically result when formed at room temperature [258]. Defectivity within oxide layers can lead to subgap leakage through the Josephson junction, predicted to be due to interface states [241],[259] at regions of lower density [196] near the metal / oxide interfaces. Interestingly, experiments conducted on Al / AlO$_x$ junctions with either an Al or Cu top metal layer suggest that increased subgap current may not be due to 'pinholes' in the oxide but rather diffusive Andreev transport [260].

Another common method of Josephson junction formation involves direct deposition and subtractive etching of the metallic and dielectric layers. These types of tunnel barrier schemes allow for a greater flexibility in materials combinations and deposition techniques, not only in the top and bottom metal electrodes but also the insulating layer [102],[261]. For example, epitaxial Re films were grown on sapphire substrates, followed by an $AlO_x$ layer that can be heteroepitaxially aligned with the underlying Re after suitable annealing, and completed with a polycrystalline Al film [261]. However, the relatively large roughness associated with the bottom Re electrode layer necessitated a multilayer approach combining Re and Ti layers, leading to improved subgap resistance and Q values on the order of 40,000 [102],[184]. In addition, AlN insulating layers within Josephson junctions have been investigated using either TiN [262] or NbN electrodes [244] on MgO substrates, though the buffering of Si substrates with TiN films resulted in substantially increased quality factors in the corresponding capacitvely shunted flux qubits [243].

Increased subgap leakage in Nb / Al / $AlO_x$ / Nb Josephson junctions has been reported with decreasing junction area, while comparable trilayer builds formed through shadow evaporation exhibited little dependence [263] suggesting that aluminum oxide damage induced by the etching process could be responsible [264]. However, differences in the residual stress state of the Nb metallization have also been used to explain degradation [265] and nonuniformity in critical current [266] of such junctions. As discussed in Section 4.3, the processing methods used to modulate the Nb stress in sputter deposited layers, involving pressure and power, can also impact the resulting microstructure and composition within the films so that the $AlO_x$ tunnel barrier properties cannot be unequivocally attributed to residual stress. In fact, reversible changes observed in $J_c$ as a function of hydrogen concentration, observed in similar junctions, has also been proposed [267] where increasing H content at the $AlO_x$ / Nb interface may increase its potential barrier height [268].

### 5.3 Characterization

Methods to assess the properties of Josephson junctions in superconducting qubits include approaches that either infer defectivity during operation or those that probe the electrical or structural characteristics through analytical techniques. In the former case, a comparative analysis using qubit spectroscopy performed on flux tunable, phase qubits demonstrated that avoided level crossings, a signature of unwanted coupling between the transition frequency and TLS, can be reduced approximately 80% by incorporating an epitaxial AlO$_x$ tunnel barrier rather than an amorphous one [261]. As described in Section 2.4, the combination of applied bias or substrate strain to such measurements allows one to indirectly determine the population of TLS within a single flux qubit; those insensitive to the applied electric field (approximately 40%) were assumed to reside within the Josephson junctions and the remainder were located on the various surfaces of the device [135],[137].

Direct characterization of tunnel barriers has often relied on cross-sectional transmission electron microscopy (TEM) to ascertain microstructural information at a nanometer spatial resolution. The degree of crystallinity, thickness and atomic coordination can be assessed through a combination of high-resolution TEM and analytical methods, such as electron energy loss spectroscopy (EELS). However, the rough morphology associated with most shadow evaporated Josephson junctions [269] can make such measurements challenging. In some cases, thicker films produced either through multiple deposition and oxidation steps of Al or electron beam evaporation of Al$_2$O$_3$ pellets are used as representative structures, revealing these layers share similar properties with those conventionally fabricated in that they are amorphous and oxygen deficient regardless of the deposition approach [270]. Variation in tunnel barrier thickness is readily observed due to the underlying Al microstructure, where regions above grain boundaries exhibit thicker $AlO_x$ [271]. Investigation of the near-edge features of the EELS spectra in concert with a pair distribution function analysis of nanobeam diffraction data indicate tetrahedrally coordinated Al atoms in the barrier interior, consistent with amorphous aluminum oxide, but less coordination near the interface indicative of oxygen vacancies [259]. Growth of Al films on Si (111) substrates revealed that Al morphology, which dictates variability in the overlying $AlO_x$ layer, is strongly dependent on the substrate treatment prior to metallic deposition that can result in aligned Al (111) grains, though ones which possess high dislocation densities due to the large lattice mismatch [272].

Surface-based, scanning techniques can be utilized to investigate barrier properties, both after formation as well as in-situ during growth. For example, scanning tunneling spectroscopy analysis of ALD $Al_2O_3$ and native aluminum oxide layers delineated differences between their potential barrier heights, where ALD films exhibited a value of approximately 1.42 eV in contrast to those found on thermally oxidized $AlO_x$ (0.67 eV) [273]. Used in combination with ballistic electron emission microscopy (BEEM), studies revealed inhomogeneous barrier properties that can vary from grain to grain in thin $AlO_x$ films [274]. In addition, low energy, electronic states present during the initial stages of $AlO_x$ growth were deemed responsible for increased subgap leakage due to variations in the local atomic structure rather than physical 'pinholes' of metallic conduction through the barrier [260]. While these measurements can provide key information relevant to the creation of the insulating layers within Josephson junctions, a clear uncertainty remains regarding modification of these surfaces after the upper metal electrode has been deposited.

From a historical perspective, the predominant amount of information gathered on Josephson junctions has been gathered using I-V characterization [159],[185]. One can derive information on the electrical properties of the insulating potential barrier (e.g: height, thickness) by analyzing data acquired in the normal state of the metallization electrodes [275], or the superconducting energy gap, critical current and subgap leakage current at temperatures below $T_c$ [226]. Note that this effective potential barrier thickness often differs from the physical thickness ascertained by X-ray reflectivity measurements [276],[277], which is often attributed to the presence of locations of enhanced conduction through the barrier layer. Because tunneling resistance scales exponentially with the thickness of the insulating layer, electrical characterization may provide a more sensitive method of ascertaining information relevant to Josephson junction performance, and is routinely used to predict superconducting qubit frequency based on the Ambegaokar-Baratoff relationship [278],[279].

**5.4 Summary**

Josephson junctions represent a critical component to many superconducting qubit implementations due to their low-loss, nonlinear behavior. Despite over five decades of research into Josephson junction formation, which has evaluated numerous metallization and tunnel barrier schemes, the most common fabrication technique for qubits still employs the oxidation of aluminum features. However, more recent work into the characterization of these junctions using complementary techniques is helping to identify sources of enhanced conductivity and subgap leakage, which can be deleterious to junction performance.

## 6. Radiation Effects

As described in the previous sections, the state of a superconducting qubit can be impacted by many methods of relaxation. Radiative energy represents the driving force for these mechanisms, whether deliberately introduced for qubit operation or from other, unwanted sources. **Fig. 13** provides a broad, pictorial representation of several elements inherent to qubits, including aspects associated with the choice of materials (yellow ellipse), design (blue ellipse) and those shared between the two, which govern their interaction with incident radiation. Although shielding and packaging schemes help to reduce its magnitude, thermal and radiative sources cannot be completely eliminated from disturbing qubits and resonators. Their manifestations generally relate to the preservation of superconductivity within the metallization, as will be detailed further in this section.

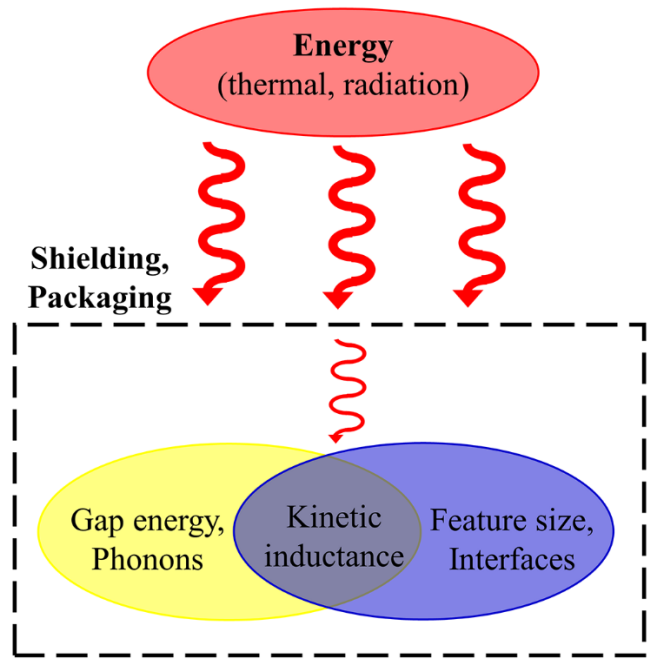

FIG. 13: Schematic of the interaction between various aspects of superconducting qubits, corresponding to materials properties (yellow), design properties (blue) and elements common to both, and incident energy, which can be reduced through shielding strategies.

## 6.1 Quasiparticles

Conventional superconductivity is realized by the combination of conduction electrons into Cooper pairs, enabled by collective ionic motion with the material [280]. This arrangement can be easily disrupted by incident energy which excites the Cooper pairs into quasiparticles. In the two-fluid model, Cooper pairs and quasiparticles represent parallel transport paths that contribute to the superconductor's complex conductivity [226], where the latter quantity can induce dissipation within quantum devices. Although the magnitude of energy must be sufficiently large to place electrons above twice the superconducting gap energy, $\Delta$, of the respective material, there is ample evidence that contributions from ambient and applied sources routinely generate quasiparticles within superconducting qubits and resonators [281],[282]. Because temperature and applied magnetic fields impact $\Delta$, these quantities can clearly increase quasiparticle creation. In addition, the overall quasiparticle density present within the metallization depends on several mechanisms that can amplify or decrease their number, such as cascades produced by burst events [283],[284] or recombination of quasiparticles back into Cooper pairs [285].

According to BCS theory, the thermal quasiparticle density, $n_{QP}$, normalized by the Cooper pair density, $n_{CP}$, represents the equilibrium quasiparticle fraction, $x_{eq}$, and, under the assumptions that $\Delta \gg k_B T$ and $hf \gg k_B T$, can be estimated by the following formula [286]:

$$x_{eq} \sim \sqrt{\frac{2\pi k_B T}{\Delta}} \, e^{-\frac{\Delta}{k_B T}} \qquad (9)$$

The activated response of $x_{eq}$ with respect to the superconducting gap energy, $\Delta$, dictates that it drop precipitously for temperatures well below the material's superconducting transition temperature, $T_c$. The qubit decay rate, $\Gamma_{qp}$, the inverse of the corresponding relaxation time, can be approximated by $\text{Re}[Y]/C_q$, where Y and $C_q$ are the complex admittance and capacitance of the qubit, respectively [287]. It can be simplified to form the following approximation based on the thermal quasiparticle fraction and non-equilibrium quasiparticle fraction, $x_{ne}$:

$$\Gamma_{qp} \sim 2f \sqrt{\frac{2\Delta}{hf}} (x_{ne} + x_{eq}) \qquad (10)$$

One would assume that the choice of metallization possessing a large $\Delta$ would mitigate the effects of radiation. However, for temperatures below approximately $T_c/10$, the athermal, non-equilibrium quasiparticle fractions remain [71].[282] and can dominate the relaxation rate. The saturation value of $\Gamma_{qp}$, if converted to an equivalent thermal energy, would correspond to approximately 160 mK [288]. As this value is unreasonably large for dilution refrigerators with base temperatures an order of magnitude smaller, it suggests other radiation mechanisms impact superconducting qubit and resonators. A variety of sources have been proposed, such as radiation from the environment [289], transmission of phonons through the metallization and the substrate [245],[284] and cosmic rays [283],[290].

Given the unavoidable existence of non-equilibrium quasiparticle production in superconducting qubits, methods to mitigate their presence have been explored. One such approach invokes quasiparticle traps, materials in contact with the qubit metallization that remain in the normal state or with a lower superconducting gap energy. Quasiparticles lose energy as they traverse into the trap material so that they possess insufficient energy to return to the original metallization [226]. However, their proximity to the qubit junctions must be chosen appropriately to avoid any deleterious effects of regions with reduced Δ [291],[292]. A similar strategy involves superconductor 'gap engineering' in the vicinity of the Josephson junctions, which reduced non-equilibrium quasiparticles in single Cooper pair transistors [293],[294], but did not appreciably affect $T_1$ values in transmon qubits [295]. Superconducting vortices, locations of concentrated magnetic field density that penetrate superconducting metallization, also provide regions of reduced Δ than the host material which can act as quasiparticle traps [296],[297]. Since their movement through the metallization can also degrade qubit and resonator performance [296],[298],[299], it becomes important to establish ways to isolate such vortices to specific sections of the superconducting material, for example by lithographically patterning slots within CPW resonators [300] or holes within the groundplane metallization [301]. However, the trapping of quasiparticles in shallow subgap states has also been proposed as a TLS that can cause relaxation [302]. In fact, increases in CPW resonator Q values at decreasing frequency could correspond to less radiation loss at longer wavelengths [63],[227].

## 6.2 Characterization

Given the prevalence of quasiparticle effects in superconducting electronics, methods to detect them in Josephson junctions [294] and resonators have been pursued for decades [158]. The effects of quasiparticles can be represented as changes to the complex admittance of superconducting resonators and qubits [286],[303], where shifts in frequency and quality factor can be directly related to quasiparticle density through the two-fluid model of superconductivity according to the Mattis-Bardeen theory [225],[227]. Such changes can be accentuated in materials that possess large kinetic inductance, which contain fewer Cooper pairs. Kinetic inductance can be amplified by reducing the resonator cross-section, either by reducing the film thickness or width [304], through which the supercurrent must pass, or by inducing disorder within the superconducting material [176],[182],[305]. In fact, hybrid resonator geometries can be used to assess the susceptibility of specific materials to such loss by employing lumped element inductors [306] or by locating them in regions of high current density [307].

The interaction of quasiparticles with Josephson junctions can also be exploited to reduce their population in flux qubits. By applying a series of sequential $\pi$ pulses on the Bloch sphere, which provides energy to quasiparticles as they tunnel across the junction, one can increase the velocity of these quasiparticles to drive them away from the vicinity of the junction. This quasiparticle pumping scheme was shown to improve quality factors in capacitively-shunted, Al flux qubits as well as minimize the observed, non-exponential $T_1$ decay rates [308]. Another method used to investigate the decay rate of qubits due to quasiparticles involved supplying a high power injection at the cavity frequency of 3D transmon qubits. The resulting decay rate of the qubits was monitored as a function of time, revealing that the effects of quasiparticle trapping and recombination varied greatly depending on the geometry of the capacitor pads [297]. Flux vortices residing in these pads were suggested to trap quasiparticles within the Al metallization, leading to a faster recovery after the energy injection in designs with short leads between the

Josephson junction and pad, while capacitors that possessed long leads exhibited a slower return to the nominal $T_1$ values, in a form indicative of quasiparticle recombination.

In qubits that possess a sufficiently low $E_J/E_C$ ratio, so that their change in transition frequency due to offset charge is measurable, one can probe quasiparticle tunneling through the Josephson junction by measuring shifts in the charge parity [309]. The sequence of a π/2 pulse, which places the qubit state on the equator of the Bloch sphere (see **Fig. 2b**), followed by an idle time equivalent to half of the difference in transition frequencies by ±e offset charge, and then a second π/2 pulse about the in-plane axis orthogonal to the first, will expose any difference due to quasiparticle tunneling when the qubit is mapped back to its basis states. Such charge parity shifts can lead to discrete fluctuations, similar in form to random telegraph noise, that can be represented by an average quasiparticle tunneling time [288],[309]. Note that these quasiparticles need not induce qubit relaxation but can simply generate a parity switch of the particular state, leading to increased dephasing [309] (see Section 7.1). Processes emanating from photons with energies greater than 2Δ have been proposed as a source for such quasiparticle tunneling events [310]. For example, measurements on offset-charge Al transmon qubits have demonstrated improved quasiparticle tunneling lifetimes with the incorporation of Eccosorb filtering placed on the input/output filtering within the Cryoperm and Al shields [73].

### 6.3 Radiation loss

While the impact of the superconducting properties on the quasiparticle fraction generated within qubits and resonators has already been discussed (Section 6.1), quasiparticle effects can also exhibit a strong dependence on the design elements of the metallization. These effects can be modulated through the incorporation of phonon traps in the groundplane [305],[311] or by the dimensions associated with the constituent structures [297] because acoustic mismatch between most metallic films and dielectric substrates [312] can mediate phonon transmission across their common interface, leading to longer QP lifetimes in metallization. For example, lower internal quality factors were observed in CPW resonators possessing larger centerline conductor widths [89] and in granular Al resonators with greater footprints [311]. Studies across a range of transmon qubit shunting capacitor dimensions [108] confirm that Q values saturate with increasing capacitor size (and decreasing participation), suggesting that another mechanism, such as radiation loss [313], is responsible. Transmon qubits with concentric paddle geometries have been applied to mitigate the electric dipole moment present in most conventional designs [314],[315].

Given its higher energy relative to most superconducting energy gaps, infrared radiation can lead to increased quasiparticle numbers so that more comprehensive shielding within the dilution refrigerator clearly has the ability to impact internal resonator and qubit quality factors [316]. At one extreme, encasing qubits within Eccosorb material reduced both the appearance of higher order transitions and improved $T_1$ values [317]. Al CPW resonators have been shown to be more sensitive to these sources of loss where Cryoperm shielding improved internal Q values by nearly a factor of 2 while comparable TiN resonators exhibited little impact [318]. Enhanced RF filtering along the microwave lines with Eccosorb and additional Al shielding surrounding Al 3D offset-charge transmon qubits led to a near doubling of quality factor and an increase in average quasiparticle tunneling time by two orders of magnitude [73],[288]. Measurements performed on Nb transmons showed that, while quasiparticle tunneling rates were highly sensitive to both the degree of shielding as well as the shunting capacitor size and shape, their relaxation times were not significantly impacted until the mixing chamber shield had been removed [70]. **Fig. 14** illustrates the extent of variation in quasiparticle tunneling rates measured as a function of capacitor design in transmon qubits, but also that their corresponding median relaxation rates ($1/T_1$) were 1 to 2 orders of magnitude greater so that, with proper shielding, quasiparticle tunneling does not currently dictate qubit relaxation [71]. Modeling of spurious antenna modes that can be generated within transmon qubit designs at frequencies much larger than that associated with their $|0\rangle$-$|1\rangle$ transition have been conducted, suggesting that quasiparticle creation from radiation with these energies can produce an effective blackbody temperature of approximately 300 mK [165].

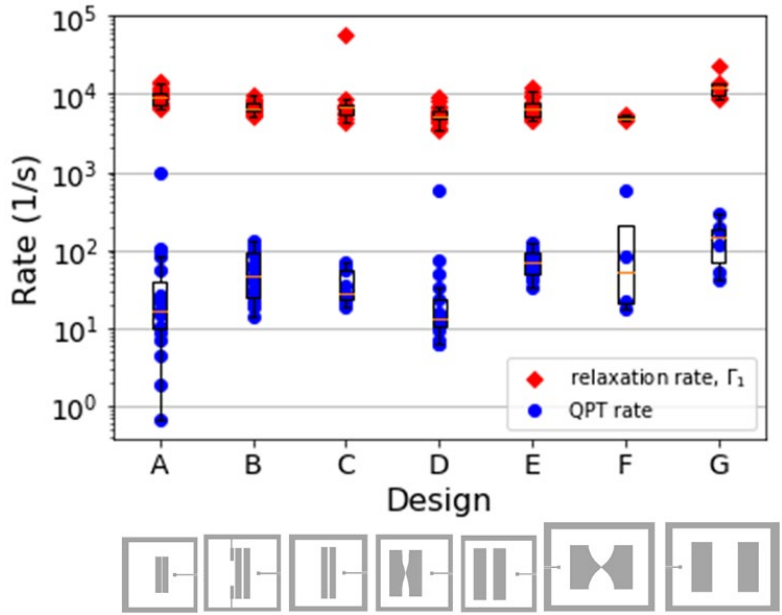

FIG. 14: Comparison of quasiparticle tunneling rates and relaxation rates ($\Gamma_1 = 1/T_1$) as a function of capacitor pad metallization design (shown below) for Nb-based, transmon qubits. Courtesy Kurter et al. [71].

The impact of ionizing radiation on resonator and qubit performance has been raised as a significant concern, where the deposited energy can generate electron-hole pairs that can decay into phonons within the metallization or underlying substrate. These high-energy cascades have the potential to create bursts of quasiparticles [305] and correlated quasiparticle relaxation in qubits on a common substrate [284],[290]. One challenge is identifying the source of such events, from impurities within the terrestrial environment to cosmic rays [289], and motivated studies that surround dilution refrigerators with lead shielding or place them within underground facilities [283]. However, mitigation strategies can also be employed at the substrate level. For example, measurements on high kinetic inductance resonators exhibited an increase in internal quality factor based on increasing the area fraction of fill features comprising the groundplane [311] or through the use of a backside groundplane [319], which has also been demonstrated to improve transmon qubit performance [313].

It is interesting that recent measurements conducted on superconducting islands, which exhibited an exponential decrease in Cooper pair breaking events over a time scale of hundreds of days at operating temperatures, suggests that sources that are effectively constant on such timescales, such as environmental radioactivity or cosmic rays, may not be the dominant mechanism [320]. Such observations hint at the possibility that thermalization issues emanating from many potential places in the qubit environment, from dielectric materials on the chiplet and packaging [303] to ancillary electronic equipment within the dilution refrigerators, still need to be addressed.

### 6.4 Summary

The presence of radiation (both intentional and unwanted) can readily generate quasiparticles within superconducting metallization. Their distribution within qubits and resonators is dictated by a complex interplay of trapping, recombination and phonon transmission processes within the constituent materials. While proper design of capacitor paddle and groundplane geometry has yielded improvements in quality factors, recent characterization of the effects of shielding and thermalization have pointed to architectures in which tunneling of quaisparticles through Josephson junctions does not appear to dictate qubit relaxation rates.

## 7. Decoherence effects

### 7.1 Dephasing

The preceding sections have primarily focused on relaxation within superconducting qubits. As shown in **Fig. 2b**, decoherence is another key parameter that describes the ability of a qubit to retain its state. Dephasing represents the shift in phase of the qubit, which refers to unwanted rotation on the Bloch sphere about the $|0\rangle$-$|1\rangle$ axis. While $T_1$ is governed by coupling to mechanisms at an energy similar to that of the transition frequency, dephasing can be impacted by a much broader spectrum. For example, motion of the qubit phase is influenced by fluctuations in the applied fields and currents in the vicinity of the qubit. Although several parameters have been used to define decoherence times, based on the measurement protocol ($T_2^E$, $T_2^*$, $T_{CPMG}$) [50], the largest value of $T_2$ achievable in a qubit according to Eq. 1, in the limit of infinite dephasing time, $T_\varphi$, would correspond to $2T_1$, where the qubit is described as '$T_1$-limited'. The incorporation of shunting capacitance, enabling high $E_J/E_C$ ratios in transmon qubits that exponentially reduces their sensitivity to charge noise, is one method to achieve this condition [321]. Flux qubits, [40],[322] where magnetic fields are deliberately introduced to tune the supercurrent through a loop of multiple Josephson junctions to maximize quality factors, will clearly be sensitive to variations in the applied flux. Coupling to flux bias line generally leads to shorter $T_1$ and $T_2$ times in these qubits over fixed frequency transmons, which are less sensitive to this decoherence mechanism [315],[323]. In fact, it has been shown that reduced tunability in flux qubits results in less sensitivity to flux instabilities [324].

Thermal photons within the qubit environment can induce significant qubit dephasing, placing more stringent conditions on the isolation of the qubit from incident and applied sources of energy. The presence of quasiparticles that tunnel across the Josephson junction can lead to reductions in $T_2$ and dephasing [325]. For example, a slight increase in the dephasing times of a transmon qubit was observed when its three-dimensional, readout cavity was filled with superfluid He [326], resulting in a slightly lower temperature but without a measurable difference in $T_1$. As will be discussed in greater detail in the following section, one source of material contributions to dephasing is defects or impurities that possess a net magnetic moment, as these can cause subgap states within the metallization, which when populated by quasiparticles can limit quality factors [327].

**7.2 Noise**

While dephasing mechanisms are only relevant to qubits, noise can be generated in the power spectral density of both superconducting resonators and qubits. This noise often scales inversely with frequency, denoted as '1/$f$' noise [328], though its form more closely resembles 1/$f^\alpha$, where the exponent α ranges from 0.6 to 1.0 [329]. Fluctuations in the critical current across Josephson junctions, charge and flux, all sources of noise within SQUID's, were proposed to be generated by the same mechanism of the magnetic moments associated with electrons which fill and deplete charge traps [330]. These traps can exist on the surfaces of oxidized metallization [331]-[333] or TLS located within the oxide barrier of the Josephson junctions [75],[86],[334].

Some of the same candidate species associated with TLS (Section 2.2) have also been associated with flux noise. Localized moments generated by MIGS near electrode-dielectric interfaces have been postulated as a source of 1/$f$ noise [329]. DFT calculations suggest extrinsic defects, such as OH [97] and $O_2$ molecules absorbed on sapphire substrates [335], where the latter quantity exhibited spin polarization when introduced on the top surface of Al and Nb films [336]. Physiabsorbed, atomic hydrogen has also been identified through both modeling [337] and electron spin resonance (ESR) characterization [338] of NbN CPW resonators to be responsible for generating flux noise. In an effort to mitigate such effects, 300 $^0$C annealing of NbN resonators led to a reduction in charge noise induced by surface spins from physisorbed H and g = 2 free radicals [339]. Similarly, backfilling of hermetic cells containing Nb SQUID's with $NH_3$ gas, which is proposed to bind to sites on the metallization surfaces preferentially over molecular oxygen, decreased the measured 1/$f$ flux noise [336].

## 7.3 Summary

The decoherence times of superconducting qubits can be impacted by a wider variety of mechanisms than their corresponding relaxation times, depending on the type of qubit (fixed frequency vs. flux tunable) as well as its thermalization. However, TLS extrinsic to the materials comprising the resonators and qubits may dictate their level of noise, so that its reduction can be accomplished by the application of appropriate surface treatments.

## 8. Conclusions and Future Outlook

Since 1999, with the successful demonstration of a superconducting qubit, quantum computing has evolved into a platform capable of conducting complex calculations on the path toward providing an advantage over classical paradigms. During its first decade of development, new qubit designs were created to limit their sensitivity to charge noise, and continuum descriptions of dielectric loss provided a means to assess the impact of insulating layers on qubit performance. The following decade extended analyses of dielectric loss to arrive at designs with less participation, while also identifying the undesired coupling of qubits to discrete TLS, the recognition of new sources of quasiparticle-induced relaxation and noise, and the realization of multi-qubit architectures. In the coming years, quantum computers will grow to encompass thousands of qubits while exploring ways to establish more uniform operating characteristics, relying not only on materials refinements but advances in the cryogenic infrastructure.

Towards these goals, there exist many open questions with respect to materials issues that deserve increased focus. Here are a few:

- The extreme sensitivity of TLS to temperature requires a precise understanding of the dielectric properties of materials at temperatures pertinent to the operating conditions of superconducting qubits (100 mK and below). This learning will continue to rely on experimental characterization to extract loss tangents of key dielectric materials. Interestingly, values for $AlO_x$ tunnel barriers from measurements of merged element transmons [340] and Cooper pair boxes [55] suggest much smaller values (below $10^{-6}$) than those typically reported for $AlO_x$ films ($\sim 10^{-3}$). Such uncertainty underscores the importance of considering the density and distribution of TLS [136] within constituent elements of the qubit; is there a threshold size [341] below which continuum descriptions of TLS behavior are no longer relevant?

- The promise of epitaxial qubit and resonator configurations is that they offer clear advantages with respect to mitigating microstructural defectivity. To date, the best quality factors have been observed in polycrystalline metallization with amorphous tunnel barriers. Are extrinsic effects due to fabrication damage still dictating performance, or are there issues with respect to plastic relaxation, such as misfit dislocations present in many heteroepitaxial configurations, that need to be addressed?

- Significant gains have been achieved in improving relaxation times over the past twenty years due to the identification of key species responsible for dielectric loss. The constraints on error rates necessary to enable fault tolerance (approximately $10^{-4}$) [342] suggests that relaxation times must be in the range of milliseconds. What issues represent the next tier of mechanisms that could limit these values? For example, piezoelectric effects induced by breaking the symmetry of centrosymmetric substrates by introducing surfaces is predicted to generate a finite loss tangent [343]. When will such effects be relevant to increasing qubit quality factors and what characterization techniques are best suited to evaluate them?

- While the primary emphasis on qubit fabrication remains increasing quality factor values, the level of non-uniformity observed in their performance represents a critical issue that will limit consistent operation of quantum computing builds, particularly in larger systems. For example, recent measurements of variation within transmon qubits reported a 3X range between minimum and maximum $T_1$ values during a span of less than three hours [73]. Some reasons for this level of variability, such as intermittent coupling to TLS [133],[136],[344] or quasiparticle bursts [284],[289],[290] have been presented. There also exists variation due to processing: for example, the spread in Josephson junction frequencies which can be reduced by treatments both during [251] and after fabrication [279],[345]. What other manufacturing aspects can be optimized to produce more uniform behavior?

- All strategies for the scaling of superconducting qubits face challenges with respect to maintaining connectivity while mitigating the generation of modes and unwanted coupling within larger packaging geometries [346]. This 'tyranny of interconnects' [347] requires the breaking of the two-dimensional geometry of a single substrate plane by using multilevel wiring routes [47] and superconducting through-silicon vias [172] to enable communication and suppressing spurious modes within groundplanes and substrates [348] as their dimensions increase. Progress has been made in enabling these approaches, for example, through indium bump bonding [233] and interposer chip arrangements [323]. However, such strategies may still be insufficient without the incorporation of larger dilution refrigerators or linking of multiple cryogenic systems [349], again placing a high importance on robust shielding and isolation of qubits [6] from non-equilibrium quasiparticle sources.

## Acknowledgments

The author would like to thank a number of his colleagues: Ken Rodbell, Martin Sandberg, Ryan Gordon, Hanhee Paik, Vivek Adiga, Karthik Balakrishnan, Ben Wymore, Santino Carnevale, Jerry Tersoff, Cihan Kurter, Oliver Dial and Matthias Steffen, for many rewarding discussions and insightful suggestions.